\documentclass[a4paper]{article}

\usepackage{import}
\usepackage{graphicx, subcaption, ragged2e}
\graphicspath{{./logos/}{./figures/}}

\usepackage{amsmath} 
\usepackage{amssymb} 
\usepackage{etoolbox}
\usepackage{mathtools}
\usepackage{commath} 

\usepackage{dsfont}

\usepackage{graphicx}
\usepackage{subcaption}
\usepackage[utf8]{inputenc} 
\usepackage[style=alphabetic]{biblatex}
\DeclareSourcemap{
  \maps[datatype=bibtex]{
    \map{
      \step[fieldsource=eprinttype, final]
      \step[fieldset=urldate, null]
      \step[fieldset=url, null]
    }
    \map{
      \pernottype{online}
      \step[fieldset=urldate, null]
      \step[fieldset=url, null]
    }
  }
}
\DeclareFieldFormat[misc]{title}{\mkbibquote{#1}}
\DeclareFieldFormat[misc]{citetitle}{\mkbibquote{#1}}

\usepackage{mdframed}
\newmdenv[
  hidealllines=true, %
  leftline=true, %
  linewidth=4pt, %
  linecolor=gray!40, %
  skipabove=\topsep, %
  skipbelow=\topsep %
]{mdquote}

\newcommand{\R}{\mathbb{R}} 

\newcommand{\Ex}{\mathbb{E}} 
\newcommand{\Exs}{\widehat{\mathbb{E}}} 
\newcommand{\Var}{\mathbb{V}} 
\newcommand{\Vars}{\widehat{\mathbb{V}}} 
\newcommand{\Cov}{\mathrm{Cov}} 
\newcommand{\Covs}{\widehat{\mathrm{Cov}}} 

\newcommand{\No}{\mathcal{N}} 
\newcommand{\MN}{\mathcal{MN}} 

\DeclareMathOperator\tr{tr}

\NewDocumentCommand{\expect}{ e{^} s o >{\SplitArgument{1}{|}}m }{%
  \operatorname{\mathbb{E}}
  \IfValueT{#1}{{\!}^{#1}}
  \IfBooleanTF{#2}{
    \expectarg*{\expectvar#4}%
  }{
    \IfNoValueTF{#3}{
      \expectarg{\expectvar#4}%
    }{
      \expectarg[#3]{\expectvar#4}%
    }%
  }%
}
\NewDocumentCommand{\expectvar}{mm}{%
  #1\IfValueT{#2}{\nonscript\;\delimsize\vert\nonscript\;#2}%
}
\DeclarePairedDelimiterX{\expectarg}[1]{[}{]}{#1}

\NewDocumentCommand{\var}{ e{^} s o >{\SplitArgument{1}{|}}m }{%
  \operatorname{\mathbb{V}}
  \IfValueT{#1}{{\!}^{#1}}
  \IfBooleanTF{#2}{
    \vararg*{\varvar#4}%
  }{
    \IfNoValueTF{#3}{
      \vararg{\varvar#4}%
    }{
      \vararg[#3]{\varvar#4}%
    }%
  }%
}
\NewDocumentCommand{\varvar}{mm}{%
  #1\IfValueT{#2}{\nonscript\;\delimsize\vert\nonscript\;#2}%
}
\DeclarePairedDelimiterX{\vararg}[1]{[}{]}{#1}

\NewDocumentCommand{\cond}{ s o >{\SplitArgument{1}{|}}m }{%
  \IfBooleanTF{#1}{
    \condarg*{\condvar#3}%
  }{
    \IfNoValueTF{#2}{
      \condarg{\condvar#3}%
    }{
      \condarg[#2]{\condvar#3}%
    }%
  }%
}
\NewDocumentCommand{\condvar}{mm}{%
  #1\IfValueT{#2}{\nonscript\;\delimsize\vert\nonscript\;#2}%
}
\DeclarePairedDelimiterX{\condarg}[1]{(}{)}{#1}

\usepackage{xparse}  

\NewDocumentCommand{\MyCondplainAuto}{mm}{%
    \left.%
    #1\vphantom{#2}%
    \,\middle\vert\,%
    #2\vphantom{#1}%
    \right.%
}

\NewDocumentCommand{\MyCondplainPlain}{mm}{%
    #1\,\vert\,#2%
}

\NewDocumentCommand{\condplain}{s >{\SplitArgument{1}{|}}m}{%
    \IfBooleanTF{#1}%
    {\MyCondplainAuto#2}
    {\MyCondplainPlain#2}
}

\newcommand{\wt}{\mathbf{w}} 

\newcommand{\Xb}{\mathbf{X}} 
\newcommand{\Sigmab}{\boldsymbol{\Sigma}} 
\newcommand{\sigmab}{\boldsymbol{\sigma}}
\newcommand{\mub}{\boldsymbol{\mu}} 
\newcommand{\betab}{\boldsymbol{\beta}}

\newcommand{\T}{\mathsf{T}} 
\newcommand{\Id}{\mathbf{I}} 
\newcommand{\uv}{\boldsymbol{1}} 

\newcommand{\Ab}{\mathbf{A}}

\newcommand{\Db}{\mathbf{D}}
\newcommand{\eb}{\mathbf{e}}
\newcommand{\Eb}{\mathbf{E}}
\newcommand{\epsilonb}{\boldsymbol{\epsilon}}
\newcommand{\varepsilonb}{\boldsymbol{\varepsilon}}
\newcommand{\Fb}{\mathbf{F}}

\newcommand{\Gb}{\mathbf{G}}
\newcommand{\Hb}{\mathbf{H}}

\newcommand{\Mb}{\mathbf{M}}

\newcommand{\rb}{\mathbf{r}}
\newcommand{\Rb}{\mathbf{R}}
\newcommand{\Sb}{\mathbf{S}}
\newcommand{\sbb}{\mathbf{s}}

\newcommand{\ub}{\mathbf{u}}
\newcommand{\Ub}{\mathbf{U}}

\newcommand{\Vb}{\mathbf{V}}

\newcommand{\Zb}{\mathbf{Z}}

\newcommand{\zerob}{\boldsymbol{0}}
\newcommand{\Gammab}{\boldsymbol{\Gamma}}

\newcommand{\Tc}{\mathcal{T}}

\DeclarePairedDelimiterX{\infdivx}[2]{(}{)}{
  #1\;\delimsize\|\;#2%
}

\newcommand{\SR}{\textnormal{SR}}

\newcommand{\IS}{\textnormal{IS}}
\newcommand{\EIS}{\textnormal{EIS}}
\newcommand{\OOS}{\textnormal{OOS}}
\newcommand{\EOOS}{\textnormal{EOOS}}

\newcommand{\PnL}{\textnormal{PnL}}

\makeatletter
\def\widebreve{\mathpalette\wide@breve}
\def\wide@breve#1#2{\sbox\z@{$#1#2$}%
     \mathop{\vbox{\m@th\ialign{##\crcr
\kern0.08em\brevefill#1{0.8\wd\z@}\crcr\noalign{\nointerlineskip}%
                    $\hss#1#2\hss$\crcr}}}\limits}
\def\brevefill#1#2{$\m@th\sbox\tw@{$#1($}%
  \hss\resizebox{#2}{\wd\tw@}{\rotatebox[origin=c]{90}{\upshape(}}\hss$}
\makeatletter

\usepackage[english]{babel} 
\usepackage{csquotes} 
\usepackage[T1]{fontenc} 
\usepackage{amsfonts} 
\usepackage[dvipsnames,svgnames,x11names,hyperref]{xcolor}

\usepackage[nottoc]{tocbibind} 

\usepackage[inline]{enumitem} 
\usepackage[normalem]{ulem} 

\usepackage{footnotebackref}

\usepackage{tablefootnote}

\numberwithin{equation}{section} 
\usepackage{titlesec} 
\definecolor{gray75}{gray}{0.75}
\titleformat{\chapter}[hang]{\Huge\bfseries}{\thechapter\hspace{20pt}\textcolor{gray75}{|}\hspace{20pt}}{0pt}{\Huge\bfseries}
\titlespacing*{\chapter}      {0pt}{0pt}{25pt}
\usepackage{geometry}
\usepackage{fancyhdr}
\usepackage{extramarks}
\fancypagestyle{plain}{}
\fancypagestyle{firststyle}
{
   \fancyhead{}
}

\makeatletter
\providecommand{\maketitle}{}
\renewcommand{\maketitle}{%
  \par
  \begingroup
    \renewcommand{\thefootnote}{\fnsymbol{footnote}}
    \renewcommand{\@makefnmark}{\hbox to \z@{$^{\@thefnmark}$\hss}}
    \long\def\@makefntext##1{%
      \parindent 1em\noindent
      \hbox to 1.8em{\hss $\m@th ^{\@thefnmark}$}##1
    }
    \thispagestyle{empty}
    \@maketitle
    \@thanks
  \endgroup
  \let\maketitle\relax
  \let\thanks\relax
}

\newcommand{\@toptitlebar}{
  \hrule height 2\p@
  \vskip 0.25in
  \vskip -\parskip%
}
\newcommand{\@bottomtitlebar}{
  \vskip 0.25in
  \vskip -\parskip
  \hrule height 2\p@
  \vskip 0.09in%
}

\usepackage[auth-lg]{authblk}
\newcommand\blfootnote[1]{%
  \begingroup
  \renewcommand\thefootnote{}\footnote{#1}%
  \addtocounter{footnote}{-1}%
  \endgroup
}

\providecommand{\@maketitle}{}
\renewcommand{\@maketitle}{%
  \vbox{%
    \hsize\textwidth
    \linewidth\hsize
    \vskip 0.1in
    \@toptitlebar
    \centering
    {\LARGE \@title\par}
    \@bottomtitlebar
    \def\And{%
      \end{tabular}\hfil\linebreak[0]\hfil%
      \begin{tabular}[t]{c}\bf\rule{\z@}{24\p@}\ignorespaces%
    }
    \def\AND{%
      \end{tabular}\hfil\linebreak[4]\hfil%
      \begin{tabular}[t]{c}\bf\rule{\z@}{24\p@}\ignorespaces%
    }
    \begin{tabular}[t]{c}\bf\rule{\z@}{24\p@}\@author\end{tabular}\center{\@date}
  }
  \setcounter{footnote}{0}
}
\newcommand{\ftype@noticebox}{8}
\newcommand{\@notice}{%
  \enlargethispage{2\baselineskip}%
  \@float{noticebox}[b]%
    \footnotesize\@noticestring%
  \end@float%
}
\makeatother

\usepackage{float}
\usepackage[section]{placeins}

\usepackage{amsthm}
\usepackage{mdframed} 
\usepackage{thmtools, thm-restate}
\usepackage{hyperref}
\definecolor{dark_grey_blue}{RGB}{0,0,200}
\definecolor{dark_grey_green}{RGB}{0,150,0}
\hypersetup{
    colorlinks,
    citecolor=dark_grey_green,
    filecolor=black,
    linkcolor=dark_grey_blue,
    urlcolor=dark_grey_blue
}
\usepackage{cleveref} 

\declaretheorem[numberwithin=section]{theorem,lemma, proposition, corollary}
\declaretheorem[style=definition,numberwithin=section]{definition, assumption}
\declaretheorem[style=remark]{remark}

\declaretheoremstyle[
headfont=\normalfont\bfseries,
notefont=\mdseries, notebraces={(}{)},
bodyfont=\normalfont,
postheadspace=0.5em,
preheadhook={\begin{mdframed}[hidealllines=true, %
  leftline=true, %
  innertopmargin=0pt, %
  innerbottommargin=0pt, %
  linewidth=4pt, %
  linecolor=gray!40, %
  innerrightmargin=0pt, %
  innertopmargin=-6pt]},
postfoothook=\end{mdframed}
]{examplestyle}

\usepackage{annotate-equations}

\usepackage{booktabs}
\usepackage{makecell}
\usepackage{tabu}

\usepackage[intoc, english]{nomencl}
\makenomenclature{}
\usepackage{tikz}
\usetikzlibrary{shapes.geometric, arrows, patterns}

\tikzstyle{rect} = [rectangle, rounded corners, 
minimum width=3cm, 
minimum height=1cm,
text centered, 
draw=black]
\tikzstyle{arrow} = [thick,->,>=stealth]
\usepackage[colorinlistoftodos,prependcaption,textsize=footnotesize]{todonotes}

\definecolor{fadedgreen}{rgb}{0.6, 0.8, 0.6}
\definecolor{lightorange}{rgb}{1, 0.7, 0.4}
\definecolor{lightred}{rgb}{1, 0.68, 0.68}

\title{In-Sample and Out-of-Sample Sharpe Ratios\\ for Linear Predictive Models}
\author[1]{Antoine Jacquier}
\author[1]{Johannes Muhle-Karbe}
\author[1, 2]{Joseph Mulligan}
\affil[1]{\normalsize Department of Mathematics, Imperial College London}
\affil[2]{\normalsize Qube Research \& Technologies}
\date{\today}

\begin{document}

\maketitle
\thispagestyle{firststyle}

\begin{abstract}
We study how much the in-sample performance of trading strategies based on linear predictive models is reduced out-of-sample due to overfitting. More specifically, we compute the in- and out-of-sample means and variances of the corresponding PnLs and use these to derive a closed-form approximation for the corresponding Sharpe ratios. We find that the out-of-sample ``replication ratio'' diminishes for complex strategies with many assets and based on many weak rather than a few strong trading signals, and increases when more training data is used. The substantial quantitative importance of these effects is illustrated with a simulation case study for commodity futures following the methodology of \citeauthor{garleanuDynamicTradingPredictable2013}~\cite{garleanuDynamicTradingPredictable2013}, and an empirical case study using the dataset compiled by~\citeauthor{goyalComprehensive2022Look2024}~\cite{goyalComprehensive2022Look2024}.\blfootnote{Corresponding Author: \href{mailto:j.mulligan21@imperial.ac.uk}{j.mulligan21@imperial.ac.uk}. Please note the disclaimer from QRT at the end of this document. The authors would like to thank Daniel Giamouridis and Paolo Guasoni for their pertinent remarks, and discussants at Imperial College London and Qube Research \& Technologies for fruitful discussions throughout the preparation of this work. The pertinent remarks of three anonymous reviewers are also gratefully acknowledged.}
\end{abstract}

\section*{Conflict of Interest Statement}
The authors have no conflicts of interest to disclose.

\section*{Data Availability Statement}
The data that support the findings of this study consists of publicly available commodity futures
prices, which are accessible from various market data providers, and the dataset compiled by Goyal, Welch and Zafirov, which is available from Amit Goyal's webpage.

\section*{Open Access Statement}
For the purpose of open access, the author(s) has applied a Creative Commons Attribution (CC BY) licence (where permitted by UKRI, `Open Government Licence' or `Creative Commons Attribution No-derivatives (CC BY-ND) licence' may be stated instead) to any Author Accepted Manuscript version `arising'.

\section*{Funding Statement}
This research was supported by Qube Research \& Technologies and the EPSRC Centre for Doctoral Training in Mathematics of Random Systems: Analysis, Modelling and Simulation (EP/S023925/1).

\section{Introduction}

A standard approach to building systematic trading strategies is to first fit some predictive model to historical training data. For example, \citeauthor{garleanuDynamicTradingPredictable2013}~\cite{garleanuDynamicTradingPredictable2013} regress price changes of commodity futures against past returns with various lookback periods. Using the forecasting model, one can then form a trading strategy, e.g., by combining forecasts for several assets using mean-variance analysis, and  ``backtest'' its performance on historical data. The strategy's ``out-of-sample'' performance on new testing data or in live trading will typically be worse than its ``in-sample'' performance on the training data for a number of reasons. But how much worse?

A number of empirical studies find that the out-of-sample performance of ETFs, anomalies and various other strategies tends to be significantly lower than the corresponding in-sample performance, with estimates ranging widely depending on the dataset~\cite{brightmanChasingPerformanceETFs2015,wieckiAllThatGlitters2016, mcleanDoesAcademicResearch2016,suhonenQuantifyingBacktestOverfitting2017,falckWhenSystematicStrategies2022}.\footnote{More generally, as in many other quantitative disciplines~\cite{ioannidisWhyMostPublished2005}, there is a controversial ongoing debate whether there is a ``replication crisis'' in financial economics~\cite{harveyCrossSectionExpectedReturns2016,jensenThereReplicationCrisis2023}.} One commonly used rule of thumb is to haircut the in-sample performance by 50\%~\cite{harveyBacktesting2015}. While such rules of thumb are useful, we seek a more systematic approach, enabling us to estimate the out-of-sample performance of different trading strategies based on their characteristics, and without having to resort to a new large-scale simulation study each time the strategy is modified.

To this end, we focus on one crucial issue in this context: \emph{overfitting}.\footnote{Another key driver of lower out-of-sample performance is \emph{multiple testing} studied by~\cite{baileyDeflatedSharpeRatio2014a, harveyBacktesting2015}. While multiple testing may lead to overfitting, we focus on overfitting from a single test perspective.} \citeauthor{kanInsampleOutofsampleSharpe2024}~\cite{kanInsampleOutofsampleSharpe2024} obtain analytical results in this spirit for the simplest setting where investment opportunities are constant, in that expected asset returns and covariances do not change over time. When these model parameters are estimated from a finite training dataset, \citeauthor{kanInsampleOutofsampleSharpe2024} derive explicit formulas for the ``replication ratio'', that is, the fraction of the in-sample Sharpe ratio that can be recovered out-of-sample.

In the present paper, we extend this analysis to a setting closer to that of \citeauthor{garleanuDynamicTradingPredictable2013} and many related papers, where trading strategies exploit several trading signals per asset and also take into account that these signals are not fully persistent but fluctuate over time. More specifically, we assume that expected returns are linear functions of iid trading signals drawn from a multivariate normal distribution. When the variance of this distribution tends to zero, the constant-parameter model of \citeauthor{kanInsampleOutofsampleSharpe2024} is recovered as a special case. Conversely, our iid expected returns correspond to the limiting case of short-lived signals in the setting of~\citeauthor{garleanuDynamicTradingPredictable2013}.

The relationship between signals and expected returns can be estimated empirically by linear regression. For portfolios corresponding to this plug-in estimate, we compute the (unconditional) in- and out-of-sample means and variances of the P\&L and derive closed-form approximations for the replication ratio, i.e., the fraction of the in-sample Sharpe ratio that can be achieved out-of-sample.

We find that as in \citeauthor{kanInsampleOutofsampleSharpe2024}~\cite{kanInsampleOutofsampleSharpe2024} the replication ratio is increasing in the size of the training dataset and the magnitude of the ``true'' Sharpe ratio of the strategy (i.e., the exact unconditional Sharpe ratio computed in the model rather than based on estimated parameters). Indeed, when the true Sharpe ratio is low the strategies are more prone to overfitting, and small increases in the true Sharpe ratio yield significant improvements in the replication ratio. Once the true Sharpe ratio is sufficiently large, the model is more robust to overfitting and the replication ratio slowly converges towards 100\%.

Conversely, the replication ratio is decreasing in the number of model parameters. 
In \citeauthor{kanInsampleOutofsampleSharpe2024}~\cite{kanInsampleOutofsampleSharpe2024} this quantity only depends on the number of assets; in our setting, it is determined both by the number of assets and by the trading signals per asset. 
The takeaway message is that complex strategies are more prone to overfitting, which needs to be balanced with sufficiently long backtesting periods. 

The practical relevance of these theoretical results is in turn explored in a simulation study that follows the implementation in~\cite{garleanuDynamicTradingPredictable2013}. To wit, we fit a linear predictive model to a set of commodity futures forecast using momentum-style signals, and in turn use this as a simulation engine. We test our approach on simulations with both Gaussian and fat-tailed returns, and for both iid and autoregressive signals. Our analytical approximations provide very good predictions in the Gaussian iid setting where they were derived, but also perform encouragingly well with autoregressive signals driven by heavy-tailed noise, confirming the practicality of our results in realistic settings. 

\begin{figure}[!ht]
    \centering
    \includegraphics{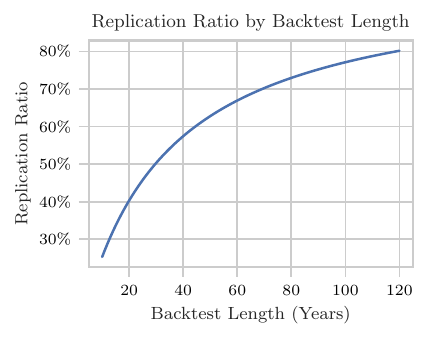}
    \caption{The replication ratio for the case study in Section~\ref{sec:pred_com_futures} with varying backtest length.}\label{fig:intro_figure}
\end{figure}

We observe that a ten-year backtest yields an expected replication ratio of just 30\% (Figure~\ref{fig:intro_figure}). This steep haircut arises from the large number of parameters in the model; it can be reduced by imposing constraints on the model, such as regularisation or a panel regression as in the empirical section of \cite{garleanuDynamicTradingPredictable2013}.

To further confirm that our results are also relevant for real data, we test the method with the comprehensive dataset compiled by~\citeauthor{goyalComprehensive2022Look2024}~\cite{goyalComprehensive2022Look2024}, which provides over 100 years of monthly signals which have been published in the literature for predicting the equity premium. This allows us to deploy the methodology developed in the present paper to subsets of the predictors and, in doing so, control both complexity of the composite strategy and backtest length. Encouragingly, we recover the key relationships from our model in the data, in that more signals lead to more overfitting, whereas longer backtests and higher strength signals mitigate the issue. 

In summary, our results suggest the following broad practical conclusions: be wary of low Sharpe ratio strategies, keep your models as simple as possible, and use the longest sensible backtest period available. While much of this advice aligns with practitioners' intuition, our results allow one to pinpoint how much of a haircut one should expect out-of-sample depending on the characteristics of the backtest, rather than broad rules of thumb. Additionally, estimating the out-of-sample haircut by simulation is often numerically slow. In contrast, our closed-form expressions permit accelerating this time-sensitive process by instantly computing the expected out-of-sample haircuts.

The remainder of this paper is organised as follows. \Cref{sec:model_setup} sets up the model and differentiates between the true and the 
sample Sharpe ratios. \Cref{sec:main_res} presents the main results and their interpretation. \Cref{sec:comp_with_lit} compares our results to the existing literature on overfitting. \Cref{sec:pred_com_futures} applies our results in a simulation case study using commodity futures, \Cref{sec:gwz} applies the method to the \citeauthor{goyalComprehensive2022Look2024} dataset, and \Cref{sec:conclusion} concludes. \Cref{sec:true_sr_deriv} and \ref{sec:deriv_n_dim} provide derivations of results used in the main text, \Cref{sec:full_result} provides the general case of the main result, and \Cref{sec:mc_analysis} provides details of the Monte Carlo analysis we use to check the approximations used.

\clearpage

\section{Model Setup}\label{sec:model_setup}
We consider a researcher who wants to fit a linear model to forecast asset returns and to trade based on these predictions. The researcher has access to some historical data observed at discrete times \(\Tc_1=\{0, \ldots, T_1\}\). We denote the returns per time step by \(\rb_{t+1}\in\R^m\) and the predictive signals by \(\sbb_t\in\R^p\) for each \(t \in \Tc_1\). The researcher assumes that the relationship between the signals and the returns follows a linear model with parameter \(\betab\in\R^{m\times p}\):
\begin{equation*}
    \rb_{t+1} = \betab\sbb_t + \epsilonb_{t+1},
\end{equation*}
where \(\epsilonb_{t+1}\in\R^m\) denotes the unexplained residual returns. We assume that the sequences $\{\sbb_t\}$ and $\{\epsilonb_{t+1}\}$ are independent of each other, and iid across time with distributions \(\mathfrak{s}\) and \(\mathfrak{e}\) respectively, which are assumed to be multivariate Gaussian:\footnote{We focus on the Gaussian case as it simplifies the exposition in many places. 
One can compute the true Sharpe ratio when the signals/noise are non-Gaussian, but we omit it here as the expression is verbose and broadly similar.}
\begin{equation*}
    \mathfrak{s}\sim\No(\mub_s,\Sigmab_s) \qquad\text{and}\qquad \mathfrak{e} \sim \No(\zerob,\Sigmab_\epsilon).
\end{equation*}
We note here that trading signals are typically assumed to be autoregressive as in \cite{garleanuDynamicTradingPredictable2013}. 
However, imposing the iid assumption as in~\cite{demarchOptimalTradingUsing2018} makes the calculations below substantially more tractable, and we demonstrate using simulation experiments in Section~\ref{sec:pred_com_futures} that the resulting formulas also provide very good approximations for autoregressive signals fitted using the methodology from the empirical part of \cite{garleanuDynamicTradingPredictable2013}.

We impose without loss of generality that the signals are centred, except for a possible intercept term, i.e., \(\mub_s = (\mu_{s,1},0,\ldots)\) where \(\mu_{s,1}=1\) if the model includes a constant (intercept) term, and \(\mu_{s,1}=0\) otherwise; 
removing the intercept simplifies many of the expressions, but we leave the expressions general as including a drift term may be desirable for some applications such as long-only equity portfolios.

Assuming \(\betab\) is known, the researcher can use the signal $\sbb_t$ at time \(t\) to accurately predict the expected returns \(\Ex_t[\rb_{t+1}]=\betab\sbb_t\) conditional on the information at time \(t\), and in turn trade a portfolio of the form \(\wt_t=\Zb\betab\sbb_t\) where \(\Zb\in\R^{m\times m}\) is some symmetric weight matrix. For example, the standard (one-period) Markowitz portfolio is given by \(\wt_t=\Sigmab^{-1}_t\mub_t\) where \(\Sigmab_t\) and \(\mub_t\) denote the covariance matrix and expected returns of the traded assets conditional on the information available at time $t$. In the present context, these are given by 
\begin{align*}
    \Ex_t[\betab\sbb_t+\epsilonb_{t+1}] &= \betab\sbb_t, \qquad \Cov_t[\betab\sbb_t+\epsilonb_{t+1}] = \Sigmab_\epsilon,
\end{align*}
and thus the one-step Markowitz optimal portfolio is of the form
$\wt_t=\Zb\betab\sbb_t$ with
\(\Zb=\Sigmab^{-1}_\epsilon\).
Somewhat less common but still relevant choices of \(\Zb\) include \(\textnormal{diag}(\Sigmab_\epsilon)^{-1}\), corresponding to an inverse variance allocation by asset, or \(\Id_m\), corresponding to a risk indifferent portfolio. 
The researcher can then test this portfolio against the returns to compute the Profit \& Loss per time step,
\begin{equation*}
    \PnL_{t} = \wt_t^\T \rb_{t+1}.
\end{equation*}

\subsection{The True Sharpe Ratio}
The Sharpe ratio typically refers to a summary statistic that compares the sample mean of a time series of returns to their sample standard deviation. In the model from the previous section, we can compare this to its ``true'' equivalent, defined as the ratio of the expected return and standard deviation of the daily P\&L, averaged over the distributions of the signal and noise terms:
\begin{equation}
    \SR := \frac{\Ex[\PnL_t]}{\sqrt{\Var[\PnL_t]}} = \frac{\Ex[\wt_t^\T\rb_{t+1}]}{\sqrt{\Var[\wt_t^\T\rb_{t+1}]}} = \frac{\Ex[(\Zb\betab\sbb_t)^\T (\betab\sbb_t + \epsilonb_{t+1})]}{\sqrt{\Var[(\Zb\betab\sbb_t)^\T (\betab\sbb_t + \epsilonb_{t+1})]}}.\label{eq:mv_sr}
\end{equation}
Using identities for quadratic expectations of Gaussian vectors, this can be computed as
\begin{equation}
    \SR = \frac{\tr(\Gb\Sigmab_s)+\mub_s^\T\Gb\mub_s}{\sqrt{2\tr((\Gb \Sigmab_s)^2) + 4\mub_s^\T\Gb\Sigmab_s \Gb \mub_s + \tr(\Fb\Sigmab_s)+\mub_s^\T\Fb\mub_s}},\label{eq:true_sr_mv}
\end{equation}
where
\begin{align*}
    \Fb \coloneqq \betab^\T\Zb\Sigmab_\epsilon\Zb\betab \quad\textnormal{and}\quad\Gb \coloneqq \betab^\T\Zb\betab.
\end{align*}
See \Cref{sec:true_sr_deriv} for the derivation of this expression. To provide some intuition for this result, suppose for simplicity that the signals include no intercept ($\mub_s=0$). Then, \eqref{eq:true_sr_mv} simplifies to
\begin{align*}
    \SR = \frac{\tr(\betab^\T\Zb\betab\Sigmab_s)}{\sqrt{2\tr((\betab^\T\Zb\betab\Sigmab_s)^2)+\tr(\betab^\T\Zb\Sigmab_\epsilon\Zb\betab\Sigmab_s)}}.
\end{align*}
In this form, clearly the expected return (numerator) increases with the Frobenius norm \(\norm{\betab}_F\) (which measures the ``predictability'' of future asset returns), however the variance (denominator) also increases. 
For a single asset or signal, the Sharpe ratio is monotonically increasing in~\(\norm{\betab}_F\), but with multiple assets or signals the relationship is no longer obvious. 
We also see that the Sharpe ratio increases as the signal-to-noise ratio rises with smaller~\(\norm{\Sigmab_\epsilon}_F\). 
Similarly to~\(\betab\), the relationship with the signal variance~\(\Sigmab_s\) is generally not monotone, but for parameters similar to our simulation case study in Section~\ref{sec:pred_com_futures} the Sharpe ratio typically increases when~\(\norm{\Sigmab_s}_F\) increases relative to~\(\norm{\Sigmab_\epsilon}_F\), again indicating an improvement in the signal-to-noise ratio of the model.

\subsection{The In-Sample Sharpe Ratio}
When a given predictive model is deployed in practice, its parameters \(\betab\) must be estimated. In the linear model we consider here, this can be done using ordinary least-squares (OLS) by stacking the returns across all time steps into the matrix \(\Rb\in\R^{m \times T_1}\) and the signals across all time steps into the matrix \(\Sb\in\R^{p \times T_1}\) and computing the OLS estimator
\begin{equation}
    \widehat{\betab} = \Rb\Sb^\T(\Sb\Sb^\T)^{-1}.\label{eq:ols_mv}
\end{equation}
Although this is the best estimator for \(\betab\) in many senses (unbiased, \(L^2\) optimal, consistent, etc.), it is also naturally overfit to the particular realisations of the signal and returns in the in-sample period, which will clearly differ in the out-of-sample period.

Moreover, using this estimate for \(\betab\) for backtesting on the same in-sample dataset naturally induces ``look-ahead bias'', as data from the full sample period has been used to estimate the model parameters which are then used to test the trading strategy on the same period.\footnote{The look-ahead bias could be avoided using a walk-forward regression as in \cite{joubertThreeTypesBacktests2024}, but this does not avoid the issue of overfitting.}

To determine the magnitude of these effects, we can also stack the residuals \(\epsilonb_{t+1}\) into the matrix \(\Eb\in\R^{m\times T_1}\) and with a substitution of the model for the returns into~\eqref{eq:ols_mv} we obtain
\begin{equation*}
    \widehat{\betab} = \betab + \Eb\Sb^\T(\Sb\Sb^\T)^{-1}.
\end{equation*}
This expression enables us to decompose the estimated regression parameter into the true regression parameter plus some sampling error. If the researcher trades the portfolio \(\widehat{\wt}_t=\Zb\widehat{\betab}\sbb_t\) using this estimate, the corresponding P\&L per time step is \(\widehat{\PnL}_t = \widehat{\wt}_t^\T\rb_{t+1}\). 
The sample average and variance in the training data set in turn are
\begin{align}
    \Exs\left[(\widehat{\PnL}_t)_{t\in \Tc_1}\right] &:= \frac{1}{T_1}\sum_{t=0}^{T_1-1} \widehat{\PnL}_{t}, \label{eq:ExpPnLHat} \\
    \Vars\left[(\widehat{\PnL}_t)_{t\in \Tc_1}\right] &:= \frac{1}{T_1-1}\sum_{t=0}^{T_1-1} \left(\widehat{\PnL}_{t}-\Exs\left[(\widehat{\PnL}_t)_{t\in\Tc_1}\right]\right)^2.\label{eq:VarPnLHat}
\end{align}
Together, these lead to the in-sample Sharpe ratio
\begin{equation*}
    \SR_\IS := \frac{\Exs\left[(\widehat{\PnL}_t)_{t\in\Tc_1}\right]}{\sqrt{\Vars\left[(\widehat{\PnL}_t)_{t\in\Tc_1}\right]}}.
\end{equation*}
The issue here is that \(\widehat{\betab}\) has been estimated using the same returns observations \(\rb_{t+1}\) which the model is then tested against, i.e.\ the model has been fit to the specific realisation of the noise in the historical period. Expanding out the sample P\&L,
\begin{align*}
    \widehat{\PnL}_t = \widehat{\wt}_t^\T\rb_{t+1} &= (\Sigmab_\epsilon^{-1}\widehat{\betab}\sbb_t)^\T(\betab\sbb_t+\epsilonb_{t+1}) \\
    &= \eqnmarkbox[fadedgreen]{truepart}{(\Sigmab_\epsilon^{-1}\betab\sbb_t)^\T\betab\sbb_t+(\Sigmab_\epsilon^{-1}\betab\sbb_t)^\T\epsilonb_{t+1}}\\
    &+\eqnmarkbox[lightorange]{misestimpart}{\left(\Sigmab_\epsilon^{-1}\Eb\Sb^\T(\Sb\Sb^\T)^{-1}\sbb_t\right)^\T \betab\sbb_t} \\
    &+ \eqnmarkbox[lightred]{overfitpart}{\left(\Sigmab_\epsilon^{-1}\Eb\Sb^\T(\Sb\Sb^\T)^{-1}\sbb_t\right)^\T \epsilonb_{t+1}}.
\end{align*}
\begin{tikzpicture}[remember picture, overlay]
    \node[anchor=west, align=left, text width=4cm] (truepart-label) at ([xshift=0.5cm]truepart.east) {Truth};
    \draw[-stealth, black] (truepart-label.west) -- ([xshift=0.1cm]truepart.east);

    \node[anchor=west, align=left, text width=4cm] (misestimpart-label) at ([xshift=0.5cm]misestimpart.east) {Misestimation};
    \draw[-stealth, black] (misestimpart-label.west) -- ([xshift=0.1cm]misestimpart.east);

    \node[anchor=west, align=left, text width=4cm] (overfitpart-label) at ([xshift=0.5cm]overfitpart.east) {Overfitting};
    \draw[-stealth, black] (overfitpart-label.west) -- ([xshift=0.1cm]overfitpart.east);
\end{tikzpicture}
We observe that by using \(\widehat{\betab}\) instead of \(\betab\) we obtain a combination of the true P\&L and some additional P\&L as a result of the estimation error in \(\betab\). In the in-sample period \(\epsilonb_{t+1}\) is an element of \(\Eb\), but in the out-of-sample period it is not (\(\Eb\) is the matrix of stacked noise realisations from the in-sample period). The impact of this is that, in expectation, the misestimation term has expected value zero in both the in- and out-of-sample period due to \(\Ex[\Eb]=\zerob\). The overfitting term however only has zero expectation in the out-of-sample period, leading to inflation in the expected return in the in-sample period, this is how the look ahead bias discussed earlier inflates the in-sample performance. However, in both periods the variance of the sample P\&L is higher than if we had used the true estimator \(\betab\), as we will see in our main result in \Cref{prop:special_case}.

\subsection{The Out-of-Sample Sharpe Ratio}

After estimating the predictive model on the training dataset, the researcher deploys the model out-of-sample at discrete trading times \(\Tc_2=\{T_1+1,\ldots,T_1+T_2+1\}\). The corresponding sample mean and variance then lead to the out-of-sample Sharpe ratio
\begin{equation*}
    \SR_\OOS := \frac{\Exs\left[(\widehat{\PnL}_t)_{t\in\Tc_2}\right]}{\sqrt{\Vars\left[(\widehat{\PnL}_t)_{t\in\Tc_2}\right]}}.
\end{equation*}
The difference, of course, is that in the future period \(\widehat{\betab}\) is not fit to the particular observations of the returns and signals, and thus the out-of-sample Sharpe ratio will typically be lower than the in-sample Sharpe ratio previously calculated. We also see from the expressions derived below that the out-of-sample Sharpe ratio will be lower than the true Sharpe ratio due to misestimation of the regression parameter causing increased volatility over trading with the true \(\betab\).

The questions therefore are: How much lower? And what can be done to mitigate this bias?

\section{Main Results}\label{sec:main_res}

The unconditional means of the in-sample and out-of-sample Sharpe ratio, \(\Ex[\SR_\IS]\) and \(\Ex[\SR_\OOS]\), cannot be computed in closed form. We can however compute the expected in-sample and out-of-sample expected return and variance. For clarity of exposition, we present here our main result for a special (but extremely common) case, taking \(\Zb=\Sigmab_\epsilon^{-1}\), \(\mu_s=\zerob\) and \(\Sigmab_s=\Id_p\), i.e., the Markowitz portfolio for a set of centred standardised signals. The impact of having to estimate the covariance matrix when constructing the portfolio is discussed in \Cref{sec:unknown_cov}. The general version of this result is reported in Appendix~\ref{sec:full_result}.
\newpage
\begin{restatable}{proposition}{propmv}\label{prop:special_case}
    The expected average in-sample P\&L and out-of-sample P\&L is given by 
    \begin{align*}
        \Ex\left[\Exs\left[\left(\widehat{\PnL}_t\right)_{t\in\Tc_1}\right]\right] &= \tr(\Gammab) + \frac{pm}{T_1}, \\
        \Ex\left[\Exs\left[\left(\widehat{\PnL}_u\right)_{u\in\Tc_2}\right]\right] &= \tr(\Gammab).
    \end{align*}
    The expected variance of the in-sample and out-of-sample P\&L is given by
    \begin{align*}
        \Ex\left[\Vars\left[\left(\widehat{\PnL}_t\right)_{t\in\Tc_1}\right]\right] &= 2\tr(\Gammab^2) + (c_1+\widetilde{c}_1)\tr(\Gammab)+c_2+\widetilde{c}_2 + \varepsilon, \\ 
        \Ex\left[\Vars\left[\left(\widehat{\PnL}_u\right)_{u\in\Tc_2}\right]\right] &= 2\tr(\Gammab^2) + c_1\tr(\Gammab)+c_2,
    \end{align*}
    where
    \begin{equation}\label{eq:gamma_constants}
        \begin{array}{rclcrcl}
        \Gammab &=& \betab^\T\Sigmab_\epsilon^{-1}\betab, \\
        c_1 &=& \displaystyle1+\frac{p+1}{T_1-p-1}, & \hspace{0.1em} &
        \tilde{c}_1 &=& \displaystyle \frac{2p+5}{T_1-p-1}+\frac{2m(p^2+p+2T_1)}{T_1(T_1-p-1)}, \\
        c_2 &=&\displaystyle\frac{m p}{T_1-p-1}, &\hspace{0.1em} &
        \tilde{c}_2 &=& \displaystyle \frac{m p \left(2 m+p+T_1+4\right)}{T_1 \left(T_1+2\right)}-\frac{2 m^2 p^2}{T_1^2 \left(T_1+2\right)}-\frac{m p}{T_1-p-1},
        \end{array}
    \end{equation}
    and 
    \begin{align*}
        \varepsilon &= \tr\left(\Cov\left[(\Sb\Sb^\T)^{-1},\sbb_t\sbb_t^\T\betab^\T\Zb\Sigmab_\epsilon\Zb\betab\sbb_t\sbb_t^\T\right]\right) + \tr\left(\Cov\left[(\Sb\Sb^\T)^{-1},\sbb_t\sbb_t^\T\betab^\T\Zb\betab\sbb_t\sbb_t^\T\right]\right) \\
        &+ \tr\left(\Cov\left[(\Sb\Sb^\T)^{-1},\sbb_t\sbb_t^\T\betab^\T\Zb\Zb\Sigmab_\epsilon\Zb\betab\sbb_t\sbb_t^\T\right]\right).
    \end{align*}
\end{restatable}

The proof relies on a combination of results on the expectations of Gaussian vectors or matrices, properties of projection matrices and matrix statistics tricks. For better readability, these lengthy computations are deferred to Appendix~\ref{sec:deriv_n_dim}. The out-of-sample expected PnL and its variance as well as the in-sample expected PnL can be computed explicitly. The in-sample variance of the PnL is also almost fully explicit, up to the \(\varepsilon\) term which does not have a closed-form expression. In practice, we approximate the in-sample variance by simply dropping this term as it is typically very small compared to the other terms. The accuracy of this approximation is verified in \Cref{sec:vareps_small}. Here, let us mention a few implications.

\begin{remark}[Expected Mean Return]
    Examining the difference between the expected in-sample and out-of-sample returns, we see that the in-sample return is inflated by \(\frac{pm}{T_1}\). 
    Thus, as the number of in-sample observations \(T_1\) increases, our in-sample P\&L will converge to our out-of-sample P\&L. Conversely, as the number of signals \(p\) increases the overfitting bias becomes arbitrarily large. Increasing the number of assets \(m\) will also lead to further inflation of the in-sample P\&L 
    (see \Cref{fig:perils_of_overfitting_replication_m_p_line_report_size}). 
\end{remark}

\begin{figure}[!ht]
    \centering
    \includegraphics{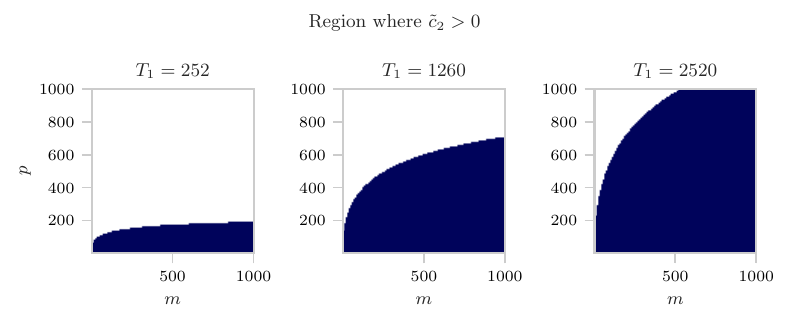}
    \caption{Values of \(m,p,T_1\) where \(\tilde{c}_2>0\).}\label{fig:tilde_c2_region}
\end{figure}

\begin{remark}[Expected Variance]\label{remark:simple_is_oos_var}
    Inspecting the expected variance, we see that \(\tilde{c}_1>0\) provided \(T_1>p+1\), which we would always expect to be the case, but whether \(\tilde{c}_2>0\) depends on the interplay between the length $T_1$ of the backtest, as well as the number $m$ and $p$ of assets and signals. For a range of backtest lengths, Figure~\ref{fig:tilde_c2_region} displays the combinations of \(m,p\) for which \(\widetilde{c}_2>0\). Generally, when \(p<m\) (with some sublinearity for large \(m\)) then \(\widetilde{c}_2>0\), and for large enough \(T_1\) then \(p\) must be much larger than \(m\) for \(\widetilde{c}_2<0\). Overfitting \(\betab\) thus causes the in-sample variance to be marginally larger than the out-of-sample variance for small \(m\), but for large \(m\) it can be significantly higher. Further, the out-of-sample variance can be much larger than the true variance, which causes the out-of-sample Sharpe to be lower than the true theoretical Sharpe ratio if \(\betab\) was known.
\end{remark}

\subsection{Accuracy of the Approximation of the In-Sample Variance}\label{sec:vareps_small}

\begin{figure}[!ht]
    \centering
    \includegraphics{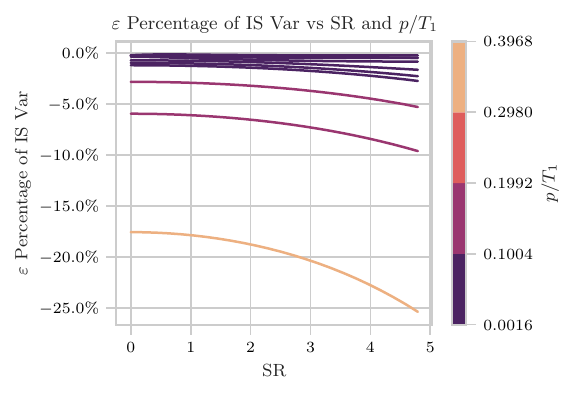}
    \caption{Magnitude of \(\sum_i\varepsilon_i\) as a percentage of the in-sample variance \(\Ex\left[\Vars\left[(\widehat{\PnL}_t)_{t\in\Tc_1}\right]\right]\).}\label{fig:vareps_pct}
\end{figure}

In our simulation study for commodity futures in Section~\ref{sec:pred_com_futures}, we find that the error term~\(\varepsilon\) from Proposition~\ref{prop:special_case} is negative but small relative to the other terms, with the estimated variance being~$3\%$ smaller than the analytical result with \(\varepsilon=0\). To support the robustness of this observation across a wide range of parameter values, we run separate simulations where the model parameters are selected randomly from distributions centred around the empirical point estimates (full details can be found in Appendix~\ref{sec:mc_analysis}). This is distinct from the case study in \Cref{sec:pred_com_futures}, where all true parameters are calibrated to that particular dataset and held fixed. We find that for \(T_1\) sufficiently larger than \(p\), \(\varepsilon\) are small, less than~\(3\%\) for $p/T_1<0.1$ as can be seen in \Cref{fig:vareps_pct}. This is equivalent to having at least ten data points per signal which in fact is a common rule of thumb~\cite[Sec.~4.4]{harrellRegressionModelingStrategies2006}. Furthermore, \(m\) has no strong effect on \(\varepsilon\) as these terms are driven by the covariance between the inverse Gram matrix \((\Sb\Sb^\T)^{-1}\) and the individual signal vectors \(\sbb_t\), which becomes negligible as the number of signals \(p\) increases. Based on these observations, we henceforth disregard~\(\varepsilon\) in our analysis.

\subsection{The Replication Ratio}\label{sec:the_repl_ratio}
In view of \Cref{prop:exp_nd} and the discussion in \Cref{sec:vareps_small} we propose the following approximations to the expected in- and out-of-sample Sharpe ratios:
\begin{equation}\label{eq:eoos_eis_approx}
    \begin{aligned}
        \Ex\left[\SR_\IS\right] \approx \SR_\EIS &:= \frac{\Ex\left[\Exs\left[(\widehat{\PnL}_t)_{t\in\Tc_1}\right]\right]}{\sqrt{\Ex\left[\Vars\left[(\widehat{\PnL}_t)_{t\in\Tc_1}\right]\right]}}, \\
        \Ex\left[\SR_\OOS\right] \approx \SR_\EOOS &:= \frac{\Ex\left[\Exs\left[(\widehat{\PnL}_t)_{t\in\Tc_2}\right]\right]}{\sqrt{\Ex\left[\Vars\left[(\widehat{\PnL}_t)_{t\in\Tc_2}\right]\right]}}.
    \end{aligned}
\end{equation}
Here we take \(\varepsilon=0\) in \(\Ex[\Vars[(\widehat{\PnL}_t)_{t\in\Tc_1}]]\) as discussed in \Cref{sec:vareps_small}. 
Moreover, we assume that the difference between taking expectations of the numerator and denominator individually, rather than jointly, is small. 

\begin{figure}[!ht]
    \centering
    \makebox[\textwidth][c]{\includegraphics[width=1.03\textwidth]{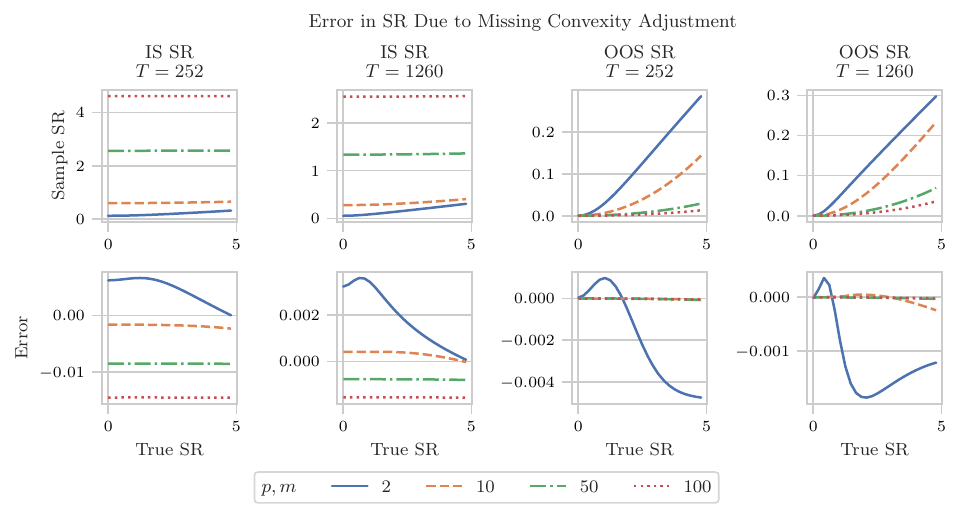}}
    \caption{Difference between the expected Sharpe Ratio \(\Ex[\SR]\) and the ratio of the expected mean and variance \(\Ex[\Exs[(\widehat{\PnL}_t)_{t\in\Tc_i}]]/\sqrt{\Ex[\Vars[(\widehat{\PnL}_t)_{t\in\Tc_i}]]}\), i.e. the error due to the missing convexity adjustment for both in-sample and out-of-sample Sharpe ratios.}\label{fig:error_convexity}
\end{figure}

We again check this by the same simulation process as \Cref{sec:vareps_small} and find that indeed the convexity adjustment is generally very small, with maximum errors of 0.01 for the in-sample Sharpe ratio and 0.005 for the out-of-sample Sharpe ratio, as can be seen in \Cref{fig:error_convexity}. Compared to the magnitude of the sample Sharpe ratios (shown in the second row), the errors due to convexity are negligible, at most 2.5\%. In these simulations we let \(p=m\) and \(T=T_1, T_2\).\footnote{Note these estimates are purely calculated via simulation, and we do not use our analytical values. This is as the error due to the missing \(\varepsilon\) terms (as studied in \Cref{sec:vareps_small}) dominates the small convexity error when \(p=100,T_1=252\), yielding no insight into the magnitude of the convexity error.}

Armed with these analytical approximations, we can study the ``replication ratio'', i.e., the fraction \(\SR_\EOOS/\SR_\EIS\) of the in-sample Sharpe ratio that is recovered out of sample. A replication ratio of \(100\%\) means that the strategy completely replicates out of sample. Conversely, a replication ratio of \(0\%\) means that it returns zero, e.g., because the signals are completely uninformative (i.e., the true parameter \(\betab\) is \(\zerob\)) and thus the expected out-of-sample Sharpe ratio is zero.

\paragraph{Univariate Case}
To gain some intuition we first evaluate the replication ratio in the simplest case with one asset (\(m=1\)) and one dynamic signal with no intercept (\(p=1\)) and where we scale~\(\sigma_s\) to be equal to~\(\sigma_\epsilon\), which is always possible as one can always scale the original signal to achieve this. In this case the formulas derived above simplify as 
\begin{align*}
    \SR &= \frac{\beta^2}{\sqrt{2\beta^4+\beta^2}}, \\
    \SR_\EIS &= \frac{\beta^2+\frac{1}{T_1}}{\sqrt{2\beta^4+\left(1+\frac{15}{T_1-2}-\frac{2}{T_1}\right)\beta^2+\frac{4}{T_1}-\frac{3}{T_1+2}-\frac{1}{T_1^2}}}, \\
    \SR_\EOOS &= \frac{\beta^2}{\sqrt{2\beta^4+\left(1+\frac{2}{T_1-2}\right)\beta^2+\frac{1}{T_1-2}}}.
\end{align*}
Whence, the replication ratio only depends on the size $T_1$ of the training dataset and the parameter~$\beta$ (or, equivalently, the true Sharpe ratio) in this case:
\begin{align}
    \frac{\SR_\EOOS}{\SR_\EIS} &= \frac{\beta ^2 \sqrt{2 \beta ^4+\left(1+\frac{15}{T_1-2}-\frac{2}{T_1}\right)\beta ^2 +\frac{4}{T_1}-\frac{3}{T_1+2}-\frac{1}{T_1^2}}}{\left(\beta ^2+\frac{1}{T_1}\right)\sqrt{2 \beta ^4+\left(1+\frac{2}{T_1-2}\right)\beta ^2+\frac{1}{T_1-2}}}.\label{eq:replication_pm_1}
\end{align}

\begin{figure}[!ht]
    \centering
    \makebox[\textwidth][c]{
    \includegraphics[width=1.05\textwidth]{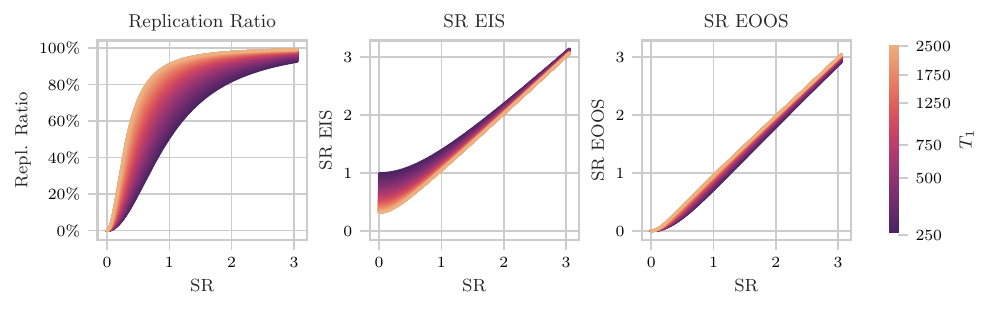}
    }
    \caption{Replication ratio \(\frac{\SR_\EOOS}{\SR_\EIS}\), \(\SR_\EIS\) and \(\SR_\EOOS\) in the 1-asset, 1-signal case as in~\eqref{eq:replication_pm_1}, with varying \(\SR,T_1\).}\label{fig:perils_of_overfitting_eoos_eis_SR_T_line_report_size}
\end{figure}

We plot a heat map of this relationship in \Cref{fig:perils_of_overfitting_eoos_eis_SR_T_line_report_size}. 
As the Sharpe ratio (equivalently~$\beta$) increases, overfitting becomes less of an issue. Indeed, as the signal-to-noise ratio becomes arbitrarily large for \(\beta\to\infty\), the replication ratio approaches 100\% ($\SR_\EOOS/\SR_\EIS\to 1$). 
Similarly, the difference between in- and out-of-sample performance also vanishes as the training data set becomes arbitrarily large ($T_1\to\infty$). 
Both these relationships are nonlinear, with large initial improvements that eventually saturate.

\paragraph{Multivariate Case}

We now turn to the case of multiple assets and signals ($m,p>1$). To illustrate some of the basic mechanisms, we look at a representative example. To wit, we consider an in-sample Sharpe ratio of \(2\) with a ten-year backtest of daily returns (\(T_1=2520\)), and assume that assets and signals are uncorrelated,  $\Sigmab_s=\Id_p$, $\Sigmab_\epsilon=\Id_m$. 
For simplicity, we also assume that the signals include no intercept ($\mub_s=\zerob$) and that investments are directly proportional to the signals without reweighing ($\Zb=\Id_m$). Finally, to achieve the desired in-sample Sharpe ratio of $2$, we let \(\betab=k\uv_{m,p}\) for a suitable constant \(k\) and the matrix \(\uv_{m,p}\in\R^{m\times p}\) of all ones. 

This is a kind of worst-case scenario for overfitting, because independent signals and returns introduce a maximal amount of freedom into the model. In this case the replication ratio becomes
\begin{equation}
    \frac{\SR_\EOOS}{\SR_\EIS} = \frac{k^2pm}{k^2pm+\frac{pm}{T_1}}\frac{\sqrt{2(k^2pm)^2 + (c_1+\widetilde{c}_1)k^2pm+c_2+\widetilde{c}_2}}{\sqrt{2(k^2pm)^2 + c_1k^2pm+c_2}},\label{eq:repl_multi_simpl}
\end{equation}
for the constants $c_i, \widetilde{c}_i$ defined  in~\eqref{eq:gamma_constants}. Whence, for independent signals and residuals, we obtain a similar expression as in the univariate case~\eqref{eq:replication_pm_1}.

\begin{figure}[!ht]
    \centering
    \includegraphics[scale=0.9]{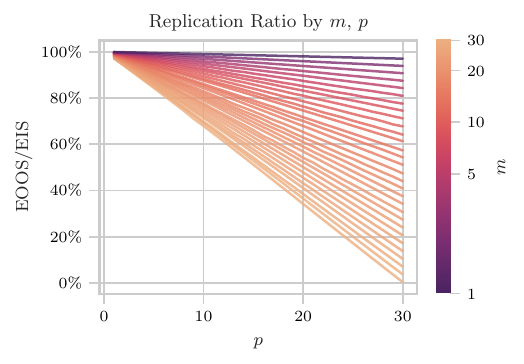}
    \caption{Replication ratio \(\frac{\SR_\EOOS}{\SR_\EIS}\) for increasing assets \(m\) and signals \(p\), and fixed in-sample Sharpe ratio \(\SR_\EIS=3\) with \(T_1=252\cdot100\). Large \(T_1\) is required to keep \(\SR_\EIS\) fixed at 3 as \(p,m\) increase, with smaller \(T_1\) we cannot always find \(\betab\) such that \(\SR_\EIS=3\).}\label{fig:perils_of_overfitting_replication_m_p_line_report_size}
\end{figure}

\Cref{fig:perils_of_overfitting_replication_m_p_line_report_size} shows that, conditional on constant in-sample Sharpe ratio \(\SR_{\EIS}=3\), the replication ratio~\eqref{eq:repl_multi_simpl} declines more quickly in \(p\) as the number of assets \(m\) increases. Why does this happen? For each additional asset or signal we increase the freedom in the model to overfit to the historical period, therefore reducing the replication ratio. 

One way to address this is to reduce the number of model parameters, e.g., by running a pooled regression that assumes that, up to some suitable normalisations, a family of common signals affects all assets in the same manner. For example, \cite{garleanuDynamicTradingPredictable2013} normalise their momentum signals for each asset to make them comparable in the cross-section and then only fit one parameter across all stocks for each of these signals, rather than many stock-specific coefficients.\footnote{This is reminiscent to the fitting of price impact models where, after normalising for each asset's volatility and trading volume, universal models also often outperform stock-specific models out of sample, as in~\cite{muhle-karbeStochasticLiquidityProxy2024} and references therein.} While it is entirely possible that the assets have differing true reactions to stimuli (for example, stocks in different sectors may respond differently to the same economic shock), in many cases this fact is overruled by the benefit that can be gained through reducing the number of parameters and avoiding overfitting. A concrete example of this is provided in \Cref{sec:pred_com_futures} where we observe that a panel regression increases the replication ratio from 71\% to 30\% for a ten-year backtest.

\begin{figure}
    \centering
    \includegraphics[scale=0.91]{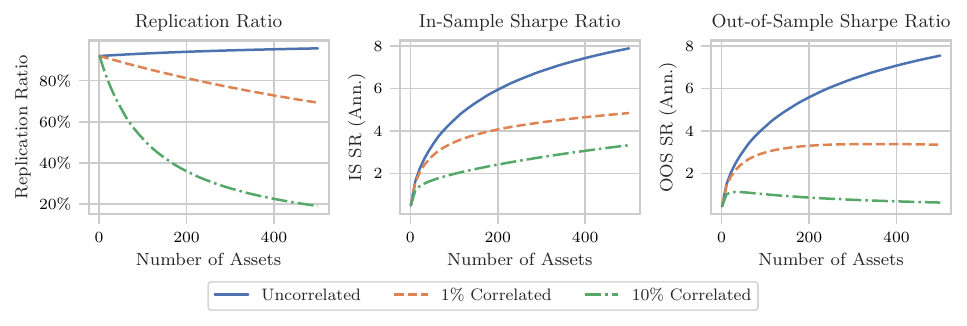}
    \caption{Replication ratio, IS Sharpe ratio and OOS Sharpe ratio as assets increase under different correlations}\label{fig:repl_oos_sr_corr_assets}
\end{figure}

Importantly, the above discussion is linked to the Sharpe ratio being held constant as the number of assets \(m\) increases. Generally, we observe that Sharpe ratios increase as we add additional uncorrelated assets to strategies, primarily due to diversification. To test when this additional performance boost outstrips the overfitting effect we consider a simple case with one signal and an increasing number of assets that the signal works equally well on. \Cref{fig:repl_oos_sr_corr_assets} shows that if these assets are completely independent, then adding additional assets results in both an increased replication ratio and out-of-sample Sharpe ratio. However, even with just 1\% average correlation between assets we see that the replication ratio  already becomes monotonically decreasing. For such a low level of correlation between the assets, the out-of-sample Sharpe ratio does initially increase as more assets are added because diversification more than compensates for overfitting. However, already with 10\% correlation this effect is reversed. In summary, adding assets generally is bad for your replication ratio. But if you can find some lowly correlated assets with consistent predictive power, it is certainly a benefit to the out-of-sample performance to include them.

\subsection{Unknown Covariances}\label{sec:unknown_cov}
In \Cref{prop:special_case} we have assumed that the true covariance matrix \(\Sigmab_\epsilon\) is known when we construct the portfolio \(\widehat{\wt}_t\). Of course, in practice we must also estimate this parameter from the observed data. This is not an easy task, and much has been written on how one can improve your estimation of the covariance matrix of returns~\cite{bunCleaningLargeCorrelation2017}. However, the corresponding estimation errors are usually of secondary importance compared to misestimation of the regression parameter \(\betab\). 

When the covariance matrix is also estimated, it is no longer feasible to obtain analytical results on the replication ratio. However, it is straightforward to assess this in a simulation study. To this end we perform a Monte Carlo simulation with $30,000$ samples where, as in \Cref{sec:vareps_small}, we select random parameters centred around the empirical point estimates from \Cref{sec:pred_com_futures} (cf.~\Cref{sec:mc_analysis} for more details) for varying numbers of signals \(p\) and assets \(m\), along with varying lookback windows \(T_1\). We form portfolios using either the known precision matrix \(\Sigmab_\epsilon^{-1}\) or the estimated precision matrix from the observed residuals \(\widehat{\Sigmab}_\epsilon^{-1}\). 

In the high-dimensional regime where the dimension \(m\) exceeds the effective sample size \(T_{\textnormal{eff}} = T_1 - 1\), the sample covariance matrix \(\widehat{\Sigmab}_\epsilon\) is rank-deficient and singular. To ensure the matrix is invertible we use a simple trace-preserving linear shrinkage estimator. We define the regularised covariance matrix as:
\begin{equation*}
    \widehat{\Sigmab}_{\textnormal{reg}} = \alpha \widehat{\Sigmab}_\epsilon + \gamma \Id_m,
\end{equation*}
\noindent where \(\Id_m\) is the identity matrix. The shrinkage intensity \(\alpha\) and the scaling factor \(\gamma\) are determined by the ratio \(q = m / T_{\textnormal{eff}}\). When \(q > 1\), we set \(\alpha = 1/q\) and \(\gamma = (1 - 1/q)\bar{\lambda}\), where \(\bar{\lambda} = \tr(\widehat{\Sigmab}_\epsilon)/m\) represents the average eigenvalue. This approach preserves the trace (total variance of the residuals) while giving us an invertible covariance matrix \(\Sigmab_{\textnormal{reg}}\). Note that when \(q<1\) we simply use the estimated covariance matrix directly.

\begin{figure}[!ht]
    \centering
    \includegraphics[scale=0.91]{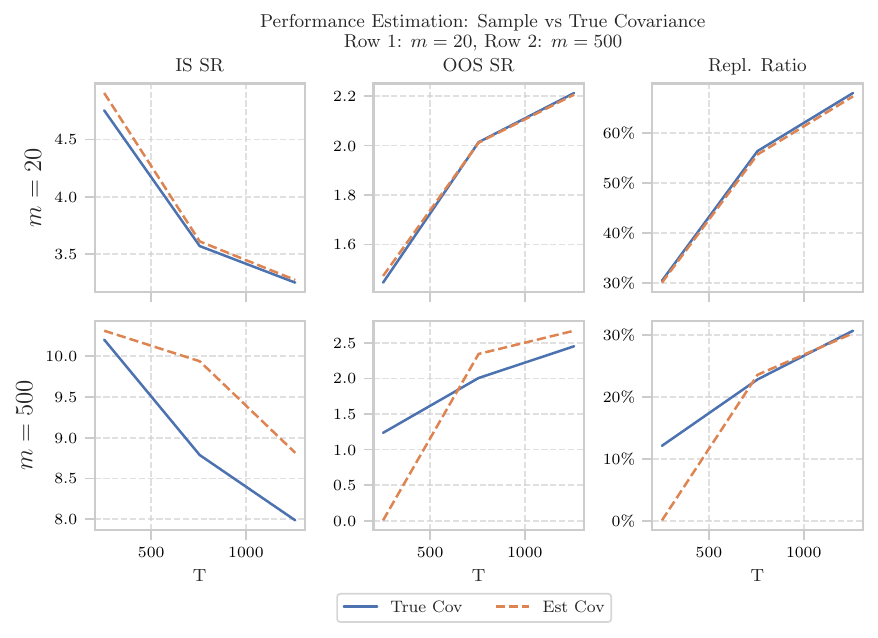}
    \caption{In- and out-of-sample Sharpe ratio and replication ratio using either the true or estimated covariance matrices \(\Sigmab_\epsilon, \widehat{\Sigmab}_\epsilon\) to form \(\widehat{\wt}_t\) in Monte Carlo simulations.}\label{fig:est_vs_true_cov}
\end{figure}

\Cref{fig:est_vs_true_cov} displays the in-sample Sharpe ratio, the out-of-sample Sharpe ratio and the replication ratio for \(m=20\) and \(m=500\). In the first case, where the asset universe is small, results using the estimated and true covariance matrices are nearly identical. In the high-dimensional setting (\(m=500\)), the in-sample Sharpe ratio is inflated further when using the estimated covariance matrix instead of the true one. Regarding the out-of-sample Sharpe ratios, we see that when the number \(T\) of time steps observed  is less than the number of assets (\(m=500\), i.e., where the covariance matrix is rank-deficient) then the out-of-sample Sharpe ratio is much worse with the estimated covariance matrix (almost zero). However, once this is resolved (\(T>m\)), the estimated portfolio actually slightly outperforms the one constructed with the true covariance matrix. Moreover, as long as \(T>m\), the replication ratio is very well aligned even when $m$ is large.

\subsection{Autoregressive Signals}\label{sec:ar_sims}

So far, we assumed for analytical tractability that signals and residuals are iid. A typical assumption 
(in~\cite{garleanuDynamicTradingPredictable2013} for example) is that instead the signal follows an AR(1) process, so that signals are 
persistent and investment opportunities in turn autocorrelated across time. Such persistent signals are not analytically tractable in our framework, but we can assess the corresponding replication ratios via simulations. We examine the simplest case with one asset and one AR(1) signal,
\begin{align*}
    r_{t+1} = \beta s_t + \epsilon_{t+1},\quad s_{t} = \phi s_{t-1} + u_{t},
\end{align*}
where $\epsilon_{t+1}\sim\No(0,1)$ and \(u_{t}\) is a centred white noise process with unit variance.\footnote{We assume that \(\abs{\phi}<1\) so that 
$(s_{t})$ is weakly stationary thus
\(\Ex[s_t]=0\) and \(\Var[s_t]=\frac{1}{1-\phi^2}\).} In this case, the true Sharpe ratio is 
\begin{equation}\label{eq:true_sr_ar_1}
    \SR = \frac{\Ex[w_t r_{t+1}]}{\sqrt{\Var[w_t r_{t+1}]}} = \frac{\beta^2\sigma_s^2}{\sqrt{2\beta^4\sigma_s^4+\beta^2\sigma^2_s}} = \frac{\beta}{\sqrt{1+2\beta^2-\phi^2}},
\end{equation}
where we have taken \(w_t=\beta s_t\), $\kappa_s$ is the kurtosis (not excess kurtosis) of the signal's noise and assumed $\beta>0$ without loss of generality. For the second equality in~\eqref{eq:true_sr_ar_1}, we have used that \(\kappa_s=3\) for \(u_t\) normally distributed, and \(\sigma_s^2=\frac{1}{1-\phi^2}\).

To understand the impact of the persistence of  AR(1) signal compared to their iid counterparts, we proceed by Monte Carlo simulations. We let \(T_2=1260\) and vary \(T_1\in\{252,1260,2520\}\), \(\phi\in\{0,0.5,0.9,0.99,0.999\}\) and \(\SR\in[0,2]\). Note that \(\phi=0.9\) corresponds to a signal half-life of~$6.5$ (daily) time-steps, whereas \(\phi=0.999\) implies a half-life of 690 time-steps. For a given signal persistence~\(\phi\) we vary the parameter~\(\beta\) to achieve the desired true Sharpe ratio in each case using~\eqref{eq:true_sr_ar_1}. 

\begin{figure}[!ht]
    \centering
    \includegraphics[scale=0.9]{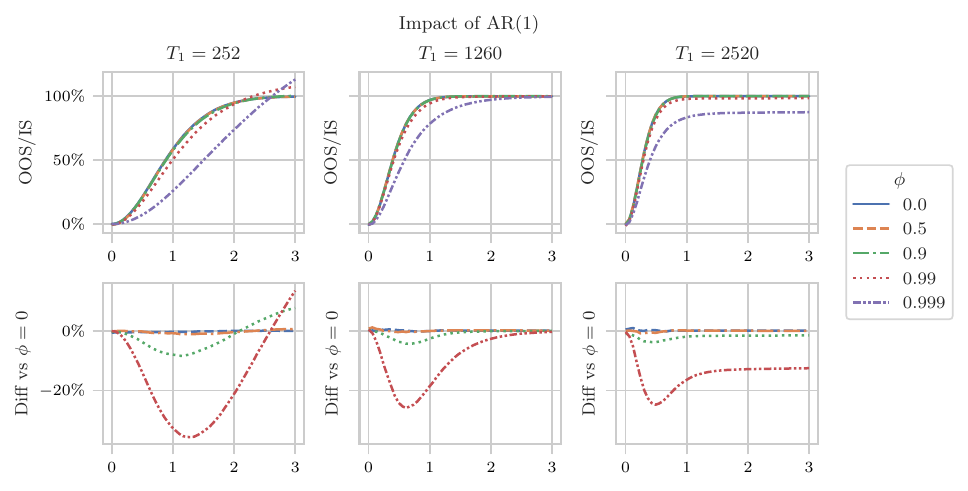}
    \caption{Impact of \(s_t\) following an AR(1) process. The top row shows the replication ratio, and the bottom row shows the simple difference of this ratio vs \(\phi=0\), corresponding to the iid case.}\label{fig:impact_ar_1}
\end{figure}

In \Cref{fig:impact_ar_1} we compare the replication ratios (and their difference to their counterparts computed in our model with iid signals) 
for different signal persistences parametrised by \(\phi\). Unless~\(\phi\) is very close to \(1\) (\(>0.99\)), the difference compared to the iid case \(\phi=0\) is surprisingly small. Thus, although our assumption of iid signals may initially seem like a strong one, it in fact has little impact on the replication ratio in practice. 

\section{Comparison with the  
Literature}\label{sec:comp_with_lit}

Most closely related to the present study is the work of~\citeauthor{kanInsampleOutofsampleSharpe2024}~\cite{kanInsampleOutofsampleSharpe2024}. They compare the expected in- and out-of-sample Sharpe ratios of the Markowitz portfolio for $m$ assets whose expected returns are estimated on a training dataset of size $T$. They find that the expected in-sample Sharpe ratio is
\begin{equation*}
    \Ex[\widehat{\theta}] = \frac{\Gamma\left(\frac{m+1}{2}\right)\Gamma\left(\frac{T-m-1}{2}\right)}{\Gamma\left(\frac{m}{2}\right)\Gamma\left(\frac{T-m}{2}\right)} {}_1F_1\left(-\frac{1}{2};\frac{m}{2};-\frac{m\theta^2}{2}\right),
\end{equation*}
where \(\Gamma(\cdot)\) is the Gamma function, \({}_1F_1(\cdot;\cdot;\cdot)\) the confluent hypergeometric function, and \(\theta=\sqrt{\mub^\T\Sigmab^{-1}\mub}\) the true Sharpe ratio of the Markowitz portfolio.
Likewise, the expected out-of-sample Sharpe ratio can also be computed in closed form as
\begin{equation*}
    \Ex[\tilde{\theta}] = \frac{\theta^2\sqrt{T}\Gamma\left(\frac{m+1}{2}\right)\Gamma\left(\frac{T-m+2}{2}\right)\Gamma\left({\frac{T}{2}}\right)}{\sqrt{2}\Gamma\left(\frac{m+2}{2}\right)\Gamma\left(\frac{T-m+1}{2}\right)\Gamma\left(\frac{T+1}{2}\right)} {}_1F_1\left(\frac{1}{2};\frac{m+2}{2};-\frac{T\theta^2}{2}\right).
\end{equation*}

\begin{figure}[!ht]
    \centering
    \includegraphics[scale=1]{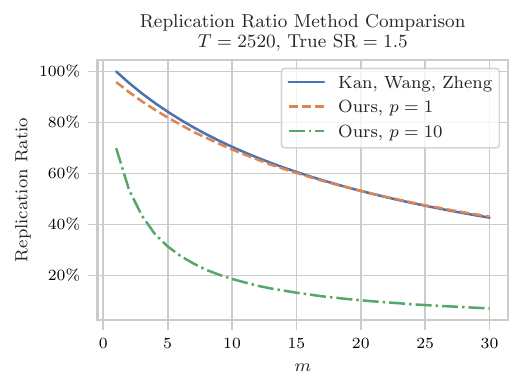}
    \caption{Comparison of the replication ratio for \citeauthor{kanInsampleOutofsampleSharpe2024} and ours with \(T=T_1=T_2=2520\) and \(\SR=1.5\) for varying \(m\). For ours we set \(p=10\) or \(p=1\).}\label{fig:kan_vs_ours_p_increasing}
\end{figure}

In the framework of \cite{garleanuDynamicTradingPredictable2013}, the setting of \cite{kanInsampleOutofsampleSharpe2024} corresponds to using a single, infinitely persistent trading signal ($\phi \to 1$) for each asset. In contrast, we consider multiple signals per asset that decay quickly ($\phi \to 0$). Incorporating multiple signals per asset allows us to cover many common trading strategies, such as momentum strategies based on several different moving averages of past returns. \Cref{fig:kan_vs_ours_p_increasing} illustrates how overfitting becomes an increasingly important concern when allowing for additional freedom in the model by considering multiple signals per asset. We fix \(T=T_1=T_2=2520\), \(\SR=1.5\), vary the number of assets \(m\) in both models and set either \(p=1\) or \(p=10\) in our model. We observe that for a single signal ($p=1$) our haircuts are similar to the ones of \cite{kanInsampleOutofsampleSharpe2024}, but for $p=10$ signals our haircut is much more severe. Note that this is conditional on a fixed true Sharpe ratio, and therefore the strength of each signal must be lower as \(p\) increases; in practice, adding signals involves a trade-off between increasing the true $\SR$ and raising the risk of overfitting.

An important common conclusion of both \cite{kanInsampleOutofsampleSharpe2024} and the present study is the non-linear dependence of the replication ratio on the true Sharpe ratio. To wit, for small values of the true Sharpe ratio, the expected out-of-sample Sharpe ratio is much lower than the expected in-sample Sharpe ratio. A takeaway from this is that when developing investment strategies, it is preferable to develop a single high Sharpe strategy than many low Sharpe strategies. This aligns with the accepted wisdom of successful practitioners, for example Peter Muller (Founder, PDT Partners) who  quipped~\cite{mullerProprietaryTradingTruth2001}:
\begin{mdquote}
    \textit{In my opinion it is far better to refine an individual strategy...than to attempt to put together lots of weaker strategies. Depth is more important than breadth for investment strategies... I would much rather have a single strategy with an expected Sharpe ratio of 2 than a strategy that has an expected Sharpe ratio of 2.5 formed by putting together five supposedly uncorrelated strategies each with an expected Sharpe ratio of 1.}
\end{mdquote}

\section{Predicting Commodity Futures}\label{sec:pred_com_futures}

To test our method we follow the approach of \citeauthor{garleanuDynamicTradingPredictable2013}~\cite{garleanuDynamicTradingPredictable2013} in the empirical application section of their seminal paper on dynamic trading. We select the commodity futures in \Cref{tbl:commodities} and collect close prices for the closest-to-expiry contracts from May 1998 to December 2023. The daily returns are computed as close-to-close percent changes in prices.

\begin{table}[!ht]
    \centering
    \begin{tabular}{l|l|l}
        Commodity & Venue & Ticker  \\ 
        \hline
        Aluminium & LME   & MAL3=LX \\
        Copper    & LME   & MCU3=LX \\
        Nickel    & LME   & MNI3=LX \\
        Zinc      & LME   & MZN3=LX \\
        Lead      & LME   & MPB3=LX \\
        Gasoil    & ICE   & LGOc1   \\
        WTI       & NYMEX & CLc1    \\
        Gold      & COMEX & GCc1    \\
        Silver    & COMEX & SIc1    \\
        Coffee    & NYBOT & KCc1    \\
        Cocoa     & NYBOT & CCc1    \\
        Sugar     & NYBOT & SBc1
    \end{tabular}
    \caption{Commodities selected for testing from the LME (London Metal Exchange), ICE (Intercontinental Exchange), NYMEX (New York Mercantile Exchange), COMEX (Commodity Exchange) and NYBOT (New York Board of Trade).}\label{tbl:commodities}
\end{table}

As in~\cite{garleanuDynamicTradingPredictable2013}, we compute the 5-day, 1-year and 5-year Sharpe ratios for each asset and use these as signals. In effect these are normalised versions of the moving averages typically used in momentum-style strategies.  We additionally transform the signals to be mean zero and unit variance. Normalizing the variances enables us to easily apply $L^2$-regularisation when fitting $\betab$ to calibrate the model we later use as a simulation engine; the centring is required per our conditions on the signals in the theoretical model. This yields a total of 36 dynamic (cross-sectional) signals, which with the addition of the asset-specific intercept yields 37 regressors per asset. 

To assess the validity of our results we need to compare our analytical approximations for the expected in-sample and out-of-sample Sharpe ratios, \(\SR_\EIS\) and \(\SR_\EOOS\), to similarly averaged observed equivalents. This requires more than the one realisation of the returns available historically. To resolve this, we simulate many realisations of a fitted model and use these to check our results. 

The first step in enabling this simulation is to estimate a model for the returns, which naturally we take as
\begin{equation*}
    \rb_{t+1} = \betab\sbb_t + \epsilonb_{t+1}.
\end{equation*}
We estimate \(\betab\in\R^{37 \times 12}\) by ridge regression with \(\gamma=0.1\). We additionally estimate the covariance of the signals \(\Sigmab_s\) and residuals \(\Sigmab_\epsilon\) from the historical data (noting that the means are zero by construction).

To simulate non-iid and non-Gaussian signals and residuals we also fit an AR(1) process to the historically observed signals,
\begin{equation*}
    \sbb'_t = \Phi \sbb'_{t-1} + \ub_{t},
\end{equation*}
where \(\sbb'_t\in\R^{36}\) denotes the non-intercept signals, \(\ub_t\in\R^{36}\) denotes the shocks to the AR(1) process, and \(\Phi \in \R^{36 \times 36}\) denotes the AR(1) parameter.

Additionally, we consider the impact of \(\epsilonb_{t+1}\) and \(\ub_t\) not being multivariate Gaussian, by fitting a \(t\)-distribution to each dimension of the residuals and shocks and using these to simulate residuals and shocks with fat tails. To induce the original covariance structure of the residuals and shocks we scale the independent samples by the Cholesky transformation of the relevant covariance matrix, \(\Sigmab_\epsilon\) or \(\Sigmab_u\).

With the fitted models we simulate $10,000$ realisations of the signals and residuals under both the Gaussian IID model, and with the AR(1) with fat tails model. All fitted parameter values from this section can be found on the authors' webpages. For context, we observe that the average correlation between signals is \(20\%\), with most signals exhibiting very low correlation, and a few exhibiting very high correlation. The average correlation between residuals is \(28\%\) and the average excess kurtosis for the residuals is~9. The diagonal of the \(\Phi\) matrix has average~0.88, and the off-diagonal has average~0.02.

As mentioned in the introduction, the haircuts we observe for these trials are quite steep, and they could be improved by reducing the amount of freedom in the model, for example by panel regression. Indeed, making the structural assumption that the normalised momentum signals have universal weights across assets reduces the number of model parameters for $m=12$ assets and $p=3\times 12+1=37$ signals each from $444$ to $3$. In simulations where the full model is used to generate the data but the panel regression is used for the portfolio construction, we see that the replication ratio is 70\%, much higher than the 38\% observed when the portfolio is constructed using the beta from the full regression. This is due to reduced overfitting, causing us to have lower in-sample Sharpe ratios (0.9 vs 8.2) but more consistent out-of-sample Sharpe ratios (0.6 vs~3.1). On the real data, when we compare panel regression with the full model we observe that the panel portfolio not only has a better replication ratio than the full model (68\% vs 9\%), but it also has a higher out-of-sample Sharpe ratio (0.5 vs 0.3), a sharp reversal given the latter's substantially better in-sample performance (panel 0.7, full 3.6). This further cements our advice to prefer simpler models wherever possible.

In addition to the study below, we conduct an additional study on the impact of using the estimated covariance matrix \(\widehat{\Sigmab}_\epsilon\) rather than the true covariance matrix \(\Sigmab_\epsilon\) to construct the portfolio, the results of which can be seen in \Cref{sec:addnl_sec5}. Broadly, we find that the impact is minimal, which affirms our previous study in \Cref{sec:unknown_cov}.

\subsection{Gaussian and iid Signals and Residuals}\label{sec:comod_fut_gauss_iid}

We first test our approach in the case where the data-generating model satisfies our key assumptions, namely that the signals and residuals are Gaussian and iid.

We simulate 13 years, with 252 trading days per year, of new signals and returns, and use 10 years as in-sample and 3 years as out-of-sample data. For each in-sample simulation we compute the OLS estimator for \(\widehat{\betab}=\Rb\Sb^\T(\Sb\Sb^\T)^{-1}\), where \(\Rb\in\R^{12 \times 252\cdot 10}\) is the stacked matrix of returns and \(\Sb\in\R^{37 \times 252 \cdot 10}\) is the stacked matrix of signals. We then form the portfolio \(\widehat{\wt}_t=\Sigmab_\epsilon^{-1}\widehat{\betab}\sbb_t\) as described in \Cref{sec:model_setup}, utilising the true covariance matrix from the historical data. (As noted in \Cref{sec:unknown_cov} estimating the covariance matrix as well only has a minor impact. This is confirmed also for this example, see \Cref{tab:all_sim_metrics_futs}). This portfolio is then tested on the in-sample and out-of-sample periods to compute the observed in-sample and out-of-sample Sharpe ratios for each simulation.

\begin{figure}[!ht]
    \centering
    \includegraphics[scale=0.9]{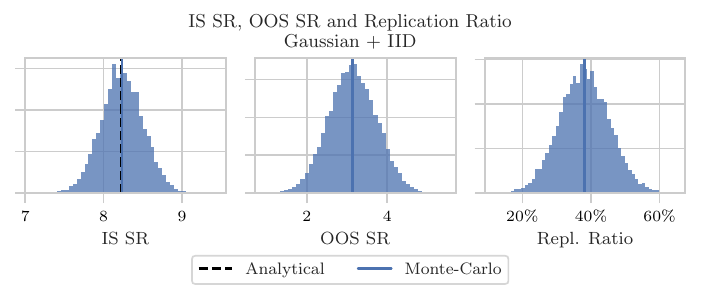}
    \caption{The expected in-sample Sharpe ratio, expected out-of-sample Sharpe ratio and replication ratio for the experiments in \Cref{sec:comod_fut_gauss_iid}.}\label{fig:comod_fut_gauss_iid}
\end{figure}

In \Cref{fig:comod_fut_gauss_iid} we see that the simulated expected in-sample and out-of-sample Sharpe ratios align perfectly with our analytical results, and so does the replication ratio. This shows that with real-world parameter values our analytical approximations are highly accurate. We note here that we have plotted the sample Sharpe ratios directly, and are thus comparing \(\Ex[\SR_\IS]\) with \(\SR_\EIS\) and similarly for the out-of-sample Sharpe ratios. For completeness, we report the average mean, variance and other metrics in \Cref{tab:all_sim_metrics_futs}.
In the next section we additionally check the robustness of our results when our assumptions are violated.

\subsection{AR(1) Signals and \(t\)-distributed Residuals and Shocks}\label{sec:comod_fut_t_ar1}

Next, we examine how our calculations hold up when the assumptions of the model are no longer true by simulating the returns and signals using the AR(1) model with fat-tail residuals and shocks.

\begin{figure}[!ht]
    \centering
    \includegraphics[scale=0.9]{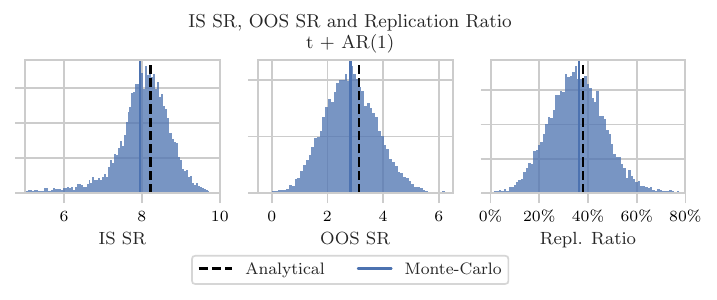}
    \caption{The expected in-sample Sharpe ratio, expected out-of-sample Sharpe ratio and replication ratio for the experiments in \Cref{sec:comod_fut_t_ar1}.}\label{fig:comod_fut_t_ar1}
\end{figure}

In \Cref{fig:comod_fut_t_ar1} we see that even when our assumptions are violated, our analytical replication ratio is still remarkably close to its simulation observed counterpart. Relaxing the Gaussian assumption to allow for fat tails causes the in-sample Sharpe ratio to decrease, but also causes the out-of-sample Sharpe ratio to decrease. Additionally, allowing the signals to be AR(1) rather than iid causes the out-of-sample Sharpe ratio to decrease, however this effect is rather small. Finally, as the out-of-sample and in-sample Sharpe ratio are affected by the non-Gaussian returns, and the AR(1) signals have a relatively small effect, the replication ratio with both is almost identical to our analytical value under the Gaussian and iid assumptions. This confirms the practical relevance of our analytical results. 

\section{Empirical Study: \citeauthor{goyalComprehensive2022Look2024} Dataset}\label{sec:gwz}

\subsubsection*{Dataset}
As another test of our approach, we use the extensive dataset compiled by~\citeauthor{goyalComprehensive2022Look2024}~\cite{goyalComprehensive2022Look2024}. The authors have constructed monthly, quarterly and annual signals which have been proposed in the academic literature to predict the equity premium of the CRSP US Index, a broad US equity index. We utilise the monthly predictors, of which there are 39 with at least some data from 1926 to 2024, almost 1200 months. These predictors are highly diverse, ranging from macroeconomic indicators, measures of investor sentiment, and variance-related signals to stock cross-sectional data, technical indicators, and commodity prices. For further details, an extensive description of each predictor is given in \cite{goyalComprehensive2022Look2024}. To pre-process these signals and the corresponding returns we wish to predict, we follow the approach of \cite{kellyVirtueComplexityReturn2024}, who also leverage this dataset, and normalise the returns by the rolling 12-month volatility and the signals by their respective expanding volatility.

\subsubsection*{Empirical Methodology}

Our analytical results are derived for the \emph{expected} replication ratio, so to check their validity we must perform some sort of averaging to mitigate the random out- or under-performance of any given signal and sample period. To achieve that on this dataset we proceed as follows. 

We repeatedly select random groups of signals of sizes between 1 and 39. We then randomly draw total lengths between 10 and 70 years and split them into (chronologically ordered) in- and out-of-sample periods with both legs at least five years long. Specifically, we perform this sampling 5,000,000 times. We constrain the sampled signals and periods to be such that each signal has full data coverage over the in- and out-of-sample period (as not all signals have data coverage for the full 1926 to 2024 period). 

For each draw, we centre the selected predictors over the entire \(T_1+T_2\) window and whiten them using a Cholesky decomposition of their sample covariance. This is useful later on when we wish to estimate \(\betab\) for our analytical values. We subsequently add a column of ones to this whitened design matrix to include an intercept in the regression model.

We then estimate the OLS parameter \(\widehat{\betab}\) using the sampled signals' in-sample data (per \Cref{eq:ols_mv}) and test this on the in-sample and out-of-sample periods using the estimated Markowitz portfolio \(\widehat{\wt}_t=\widehat{\Sigmab}_\epsilon^{-1}\widehat{\betab}\sbb_t\), where \(\widehat{\Sigmab}_\epsilon^{-1}\) is also estimated from the observed in-sample residuals. In this case we cannot know the true covariance matrix \(\Sigmab_\epsilon\), but as we are predicting just one asset (the CRSP Index), \(\widehat{\Sigmab}_\epsilon\) is simply the variance of the residuals \(\sigma_\epsilon^2\), which is relatively easy to estimate (in comparison to a large covariance matrix of many assets).

\subsubsection*{Parameter Estimation for Analytical Values}
This process yields pairs of in-sample and out-of-sample Sharpe ratios \(\SR_\IS\) and \(\SR_\OOS\). According to our analysis above, the primary drivers of their ratio \(\SR_\OOS/\SR_\IS\) should be: the number of signals \(p\), in-sample periods~\(T_1\), and true signal strength \(\betab\) for the strategy. Clearly, the number of signals \(p\) and backtest length~\(T_1\) are directly available here, but we will need to estimate \(\betab\). From our analysis throughout the paper, we've shown that estimating \(\betab\) is quite difficult. 

However, for our needs, the exact value of each entry of \(\betab\) is less important than having the correct magnitude of \(\betab\) and in turn the correct magnitude of the true Sharpe ratio (see \Cref{fig:perils_of_overfitting_eoos_eis_SR_T_line_report_size}). We found in experimentation that using the \(\widehat{\betab}\) value directly led to the analytical values being poor estimates, and therefore devised the following method to calibrate \(\betab\) as an input to the analytical computations. Specifically, rather than using \(\widehat{\betab}\) directly, we compute an implied parameter \(\betab^\star\) by minimising the difference between the observed out-of-sample Sharpe ratio \(\SR_\OOS\) and the expected out-of-sample Sharpe ratio \(\SR_\EOOS\):
\begin{align*}
    \betab^\star 
        &= \arg\min_{\betab} \left( \SR_\OOS - \SR_\EOOS(\mub_s, \Sigmab_s, \Sigmab_\epsilon, T_1, \betab) \right)^2.
\end{align*}
We perform this calibration separately for each resample. We fix the intercept coefficient within~\(\betab\) to zero (assuming we predict day-to-day price moves but not the unconditional drift). For the remaining active signals, \(\mub_s=\zerob_p\) and \(\Sigmab_s=\Id_p\) due to our whitening step. Additionally, \(\Sigmab_\epsilon\) is estimated by \(\widehat{\Sigmab}_\epsilon\) from the particular sample, and \(T_1\) is the length of the sample's in-sample period. 

For the minimisation, we take a reduced form for \(\betab\) and assume equal predictive power for all the active signals, \(\betab=(0, k, k, \ldots)\). We do this to further reduce the degrees of freedom. We use a root-finding routine to choose the value of \(k\) such that the analytical \(\SR_\EOOS\) matches the realised out-of-sample Sharpe ratio for each specific trial. Consequently, we view the resulting \(\betab^\star\) as a local calibration of the true Sharpe ratio for that particular set of signals and period.

Once we have this implied \(\betab\), we can use it to estimate the expected in- and out-of-sample mean, variance and Sharpe ratio, and expected replication ratio for that particular set of signals over that particular sampled time period. We note here that in the following analysis, all analytical values use this implied \(\betab\).

\subsubsection*{Analysis}

We recap what we now have in hand. From the process described above we have obtained many sets of observed in- and out-of-sample empirical and analytical metrics. As discussed, the analytical metrics are expectations, while the empirical metrics are simply observed data points for a particular set of signals over a particular period, which will naturally have measurement error and deviate from their (unknown) expected values. 

To resolve this, we look at averages for both the empirical and analytical metrics as a function of our identified key variables: number of signals \(p\), length of backtest \(T_1\) and true Sharpe ratio \(\SR\). By computing the averages as a function of these variables across all the samples, the effect of selecting different signals over different periods, and the finite sample measurement error, should average out, and our average empirical values should approach our average analytical values. 

\begin{figure}[!ht]
    \centering
    \includegraphics{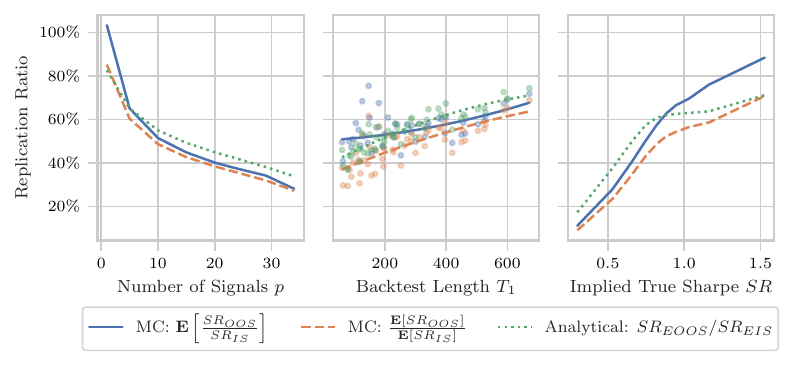}
    \caption{Empirical and implied analytical replication ratio by number of signals \(p\), backtest length \(T_1\) and implied true Sharpe ratio \(\SR\) (annualised).}\label{fig:gwz_results}
\end{figure}

The results of this procedure are displayed in \Cref{fig:gwz_results}. In each panel, we plot both the expected ratio of the out-of-sample and in-sample Sharpe ratios, and the ratio of their expectations, and the analytical replication ratio. Generally, we see that all the estimates are very close to each other. In the first panel, we observe that indeed the replication ratio decreases considerably as the number of signals \(p\) increases.

In the second panel, we plot the replication ratio as a function of \(T_1\). We found that this was not smoothly monotonic (perhaps due to insufficient samples), so we plot both the means per backtest length in our samples, and a smooth fit. We see that in general the three estimates agree, with the replication ratio increasing as the backtest length increases.

The final panel shows the replication ratio as a function of the implied true Sharpe ratio. Again, all three cases show the expected (increasing) behaviour, and the empirical value, which corresponds more closely to our analytical estimate (the ratio of the expectations) is indeed closer to our analytical values than the alternative (expectation of the ratios). In addition to the replication ratio displayed, we report the individual metrics (mean, variance and Sharpe ratio) in \Cref{sec:addnl_sec}.

\section{Conclusion}\label{sec:conclusion}

This paper derives analytical approximations for  the in-sample and out-of-sample Sharpe ratios of portfolios constructed using linear prediction models. We show that increasing either the number of signals or assets too much makes this procedure susceptible to overfitting and thereby yields wildly overestimated in-sample Sharpe ratios.

We show that low true Sharpe ratio signals are particularly vulnerable to overfitting. 
Conversely, by extending the length of the in-sample period one can reduce the overfitting risk, and can produce a higher replication ratio out of sample.

We test our results on commodity futures using momentum-style signals and find that allowing AR(1) signals and non-Normal signals/residuals does not significantly impact the validity of our results. In particular, once we match the theoretical out-of-sample Sharpe ratio to the observed value, we see that the replication ratio is primarily a function of the out-of-sample Sharpe ratio and the curves for the AR(1) signals closely matches those of iid signals.

From this analysis, it seems that the best way to minimise the potential for overfitting is to minimise the number of signals and assets that are being used for any predictive model used to trade, and utilise the largest amount of data possible. 
To conclude, we quote Nick Patterson~\cite{pattersonTalkingMachinesAI2016} on his experience at Renaissance Technologies, which is very much in line with these results:
\begin{mdquote}
    \textit{It's funny that I think the most important thing to do on data analysis is to do the simple things right. So, here's a kind of non-secret about what we did at 
    Renaissance: in my opinion, our most important statistical tool was simple regression with one target and one independent variable.}
\end{mdquote}
They might be on to something.

\clearpage
\fancyhead[L]{\slshape References}
\printbibliography[title={References}]

\clearpage
\fancyhead[L]{\firstleftmark}
\appendix
\section{Derivation of \Cref{eq:true_sr_mv}}\label{sec:true_sr_deriv}
To derive~\eqref{eq:true_sr_mv} we need to derive the  expressions for
\begin{equation*}
\Ex[(\Zb\betab\sbb_t)^\T (\betab\sbb_t + \epsilonb_{t+1})]
\qquad \textnormal{and} \qquad \Var[(\Zb\betab\sbb_t)^\T (\betab\sbb_t + \epsilonb_{t+1})].
\end{equation*}
We start with the former:
\begin{align*}
    \Ex[(\Zb\betab\sbb_t)^\T (\betab\sbb_t + \epsilonb_{t+1})] &= \Ex[\sbb_t^\T\betab^\T\Zb\betab\sbb_t + \sbb_t^\T\betab^\T\Zb\epsilonb_{t+1}], \\
    &= \Ex[\sbb_t^\T\betab^\T\Zb\betab\sbb_t], \\
    &= \tr(\Gb\Sigmab_s) + \mub_s^\T\Gb\mub_s,
\end{align*}
where \(\Gb \coloneqq \betab^\T\Zb\betab\). We have used \(\Ex[\epsilonb_{t+1}]=0\) in the second equality, and identity 5.24 from \cite{brookesMatrixReferenceManual2020} in the final equality. 

Next we tackle the variance,
\begin{align*}
    \Var[(\Zb\betab\sbb_t)^\T (\betab\sbb_t + \epsilonb_{t+1})] &= \Var[\sbb_t^\T\betab^\T\Zb\betab\sbb_t + \sbb_t^\T\betab^\T\Zb\epsilonb_{t+1}], \\
    &= \Var[\sbb_t^\T\betab^\T\Zb\betab\sbb_t] + \Var[\sbb_t^\T\betab^\T\Zb\epsilonb_{t+1}] + \Cov[\sbb_t^\T\betab^\T\Zb\betab\sbb_t, \sbb_t^\T\betab^\T\Zb\epsilonb_{t+1}], \\
    &= \Var[\sbb_t^\T\betab^\T\Zb\betab\sbb_t] + \Var[\sbb_t^\T\betab^\T\Zb\epsilonb_{t+1}], \\
    &= 2 \tr((\Gb\Sigmab_s)^2)+4\mub_s^\T\Gb\Sigmab_s\Gb\mub_s + \Var[\sbb_t^\T\betab^\T\Zb\epsilonb_{t+1}], \\
    &= 2 \tr((\Gb\Sigmab_s)^2)+4\mub_s^\T\Gb\Sigmab_s\Gb\mub_s + \Ex[\sbb_t^\T\betab^\T\Zb\epsilonb_{t+1}\sbb_t^\T\betab^\T\Zb\epsilonb_{t+1}]-\Ex[\sbb_t^\T\betab^\T\Zb\epsilonb_{t+1}]^2, \\
    &= 2 \tr((\Gb\Sigmab_s)^2)+4\mub_s^\T\Gb\Sigmab_s\Gb\mub_s + \Ex[\sbb_t^\T\betab^\T\Zb\epsilonb_{t+1}\sbb_t^\T\betab^\T\Zb\epsilonb_{t+1}], \\
    &= 2 \tr((\Gb\Sigmab_s)^2)+4\mub_s^\T\Gb\Sigmab_s\Gb\mub_s + \Ex[\sbb_t^\T\betab^\T\Zb\Sigmab_\epsilon\Zb\betab\sbb_t], \\
    &= 2 \tr((\Gb\Sigmab_s)^2)+4\mub_s^\T\Gb\Sigmab_s\Gb\mub_s + \tr(\Fb\Sigmab_s)+\mub_s^\T\Fb\mub_s,
\end{align*}
where \(\Fb \coloneqq \betab^\T\Zb\Sigmab_\epsilon\Zb\betab\). We have used \(\Ex[\epsilonb_{t+1}]=0\) to deduce \(\Cov[\sbb_t^\T\betab^\T\Zb\betab\sbb_t, \sbb_t^\T\betab^\T\Zb\epsilonb_{t+1}]=0\), identity 379 from \cite[sec. 8.2.2]{petersenMatrixCookbook2012} to compute \(\Var[\sbb_t^\T\betab^\T\Zb\betab\sbb_t]\), identity 322 from \cite[sec. 6.2.2]{petersenMatrixCookbook2012} to compute the penultimate line and identity~5.24 from \cite{brookesMatrixReferenceManual2020} in the final equality.

Thus,
\begin{equation*}
    \SR = \frac{\tr(\Gb\Sigmab_s)+\mub_s^\T\Gb\mub_s}{\sqrt{2\tr((\Gb \Sigmab_s)^2) + 4\mub_s^\T\Gb\Sigmab_s \Gb \mub_s + \tr(\Fb\Sigmab_s)+\mub_s^\T\Fb\mub_s}},
\end{equation*}
where
\begin{align*}
    \Fb \coloneqq \betab^\T\Zb\Sigmab_\epsilon\Zb\betab \quad\textnormal{and}\quad\Gb \coloneqq \betab^\T\Zb\betab.
\end{align*}

\section{General Case Result of \Cref{prop:special_case}}\label{sec:full_result}
\begin{restatable}{proposition}{propmv}\label{prop:exp_nd}
    The expected average in-sample P\&L is 
    \begin{equation*}
        \Ex\left[\Exs[(\widehat{\PnL}_t)_{t\in\Tc_1}]\right] = \tr(\Gb\Sigmab_s)+\mub_s^\T\Gb\mub_s + \frac{p}{T_1}\tr(\Zb\Sigmab_\epsilon).
    \end{equation*}
    The expected variance of the in-sample P\&L is 
    \begin{align*}
        &\Ex\left[\Vars\left[(\widehat{\PnL}_t)_{t\in\Tc_1}\right]\right] \\
        &= 2\tr\left((\Gb \Sigmab_s)^2\right) + 4\mub_s^\T \Gb \Sigmab_s \Gb \mub_s + \tr(\Fb\Sigmab_s)+\mub_s^\T\Fb\mub_s \\
        &\quad+ \frac{3}{T_1-p-1}\tr\left(2\Fb(\Sigmab_s+\mub_s\mub_s^\T)+\mub_s^\T\Fb\mub_s\cdot(\Sigmab_s+\mub_s\mub_s^\T)^{-1}(\Sigmab_s-\mub_s\mub_s^\T)+p\Fb\Sigmab_s\right) \\
        &\quad+ \tr(\Sigmab_\epsilon\Zb\Sigmab_\epsilon\Zb)\left(\frac{p \left(p+T_1+4\right)}{T_1 \left(T_1+2\right)}-\frac{2 \uv_p^\T\mub_s  \left(p-T_1-2\right) \left(p-T_1\right)}{T_1^2 \left(T_1+1\right) \left(T_1+2\right)}\right) \\
        &\quad-\tr(\Sigmab_\epsilon\Zb)^2\frac{2 \left(p-T_1\right) \left((p-2) \uv_p^\T\mub_s+T_1 \left(p-\uv_p^\T\mub_s\right)+p\right)}{T_1^2 \left(T_1+1\right) \left(T_1+2\right)}\\
        & \quad+\frac{2}{T_1-p-1}
        \tr(\Zb\Sigmab_\epsilon)\tr\left(2\Gb\left(\Sigmab_s+\mub_s\mub_s^\T\right) + \mub_s^\T\Gb\mub_s\cdot(\Sigmab_s+\mub_s\mub_s^\T)^{-1}(\Sigmab_s-\mub_s\mub_s^\T)+p\Gb\Sigmab_s\right) \\
        & \quad- \frac{2p}{T_1}
        \tr(\Zb\Sigmab_\epsilon)\left(\tr(\Gb\Sigmab_s) + \mub_s^\T\Gb\mub_s\right) \\
        & \quad+ \varepsilon_1 + \varepsilon_2 + \varepsilon_3,
    \end{align*}
    where \(\Fb = \betab^\T\Zb\Sigmab_\epsilon\Zb\betab\), \(\Gb = \betab^\T\Zb\betab\) and
    \begin{align*}
        \varepsilon_1 &= \tr\left(\Cov\left[(\Sb\Sb^\T)^{-1},\sbb_t\sbb_t^\T\betab^\T\Zb\Sigmab_\epsilon\Zb\betab\sbb_t\sbb_t^\T\right]\right), \\
        \varepsilon_2 &= \tr\left(\Cov\left[(\Sb\Sb^\T)^{-1},\sbb_t\sbb_t^\T\betab^\T\Zb\betab\sbb_t\sbb_t^\T\right]\right), \\
        \varepsilon_3 &= \tr\left(\Cov\left[(\Sb\Sb^\T)^{-1},\sbb_t\sbb_t^\T\betab^\T\Zb\Zb\Sigmab_\epsilon\Zb\betab\sbb_t\sbb_t^\T\right]\right),
    \end{align*}
    for any \(t\in\Tc_1\) by the iid property of \((\sbb_t)_{t\in\Tc_1}\). 
    
     The expected average out-of-sample P\&L is 
    \begin{equation*}
        \Ex\left[\Exs\left[(\widehat{\PnL}_t)_{t\in\Tc_2}\right]\right] = \tr(\Gb\Sigmab_s)+\mub_s^\T\Gb\mub_s,
    \end{equation*}
       The expected variance of the out-of-sample P\&L is 
    \begin{align*}
        &\Ex\left[\Vars\left[(\widehat{\PnL}_t)_{t\in\Tc_2}\right]\right] \\
        &= 2\tr\left((\Gb\Sigmab_s)^2\right) + 4\mub_s^\T \Gb \Sigmab_s \Gb \mub_s + \tr(\Fb\Sigmab_s)+\mub_s^\T\Fb\mub_s \\
        & \quad+ \mub_s^\T\left(2\Db\Sigmab_s^\T\Fb+\tr\left(\Fb\Sigmab_s^\T\right)\Db+\tr\left(\Sigmab_s^\T\Db\right)\Fb\right)\mub_s + \tr\left(\Db\Sigmab_s^\T\Fb\Sigmab_s\right)+\tr\left(\Db\Sigmab_s\right)\tr\left(\Fb\Sigmab_s^\T\right) \\
        &\quad + \tr\left((\Zb\Sigmab_\epsilon)^2\right)\left(\tr\left(\Sigmab_s^\T\Db\right) + \mub_s^\T\Db\mub_s\right),
    \end{align*}
    where \(\displaystyle
    \Db \coloneqq \frac{\left(\Sigmab_s+\mub_s\mub_s^\T\right)^{-1}}{T_1-p-1}\).
\end{restatable}
\section{Proof of \Cref{prop:exp_nd}}\label{sec:deriv_n_dim}
Recall that the classical Ordinary Least Square method yields the estimator
\begin{equation*}
    \widehat{\betab} = \betab + \Eb\Sb^\T(\Sb\Sb^\T)^{-1},
\end{equation*}
where \(\Eb\in\R^{m \times T_1}\) is the matrix of stacked residuals. This is a more convenient expression to work with than~\eqref{eq:ols_mv}, and we will use it to evaluate the expectations of the sample means, variances and covariances. 
Since \((\sbb_t)_t\) and \((\epsilonb_{t+1})_t\) are iid sequences, the expectation of the sample expectation, variance and covariance are simply the true expectation, variance and covariance respectively.

\subsection{In-sample expected average}
From the definition~\eqref{eq:ExpPnLHat}, 
the in-sample expected P\&L 
$\Ex[\Exs[(\widehat{\PnL}_t)_{t\in\Tc_1}]]$
can be evaluated as
\begin{align*}
    & \Ex\left[\Exs\left[\left((\Zb\widehat{\betab}\sbb_t)^\T(\betab\sbb_t+\epsilonb_{t+1})\right)_{t\in\Tc_1}\right]\right] \\
    & = \Ex\left[\Exs\left[\left(\{\Zb\widehat{\betab}\sbb_t\}^\T\betab\sbb_t\right)_{t\in\Tc_1}\right]\right] + \Ex\left[\Exs\left[\left(\{\Zb\widehat{\betab}\sbb_t\}^\T\epsilonb_{t+1}\right)_{t\in\Tc_1}\right]\right], \\
    & = \Ex\left[\Exs\left[\left(\left\{\Zb\left(\betab + \Eb\Sb^\T(\Sb\Sb^\T)^{-1}\right)\sbb_t\right\}^\T\betab\sbb_t\right)_{t\in\Tc_1}\right]\right] \\ 
    &+ \Ex\left[\Exs\left[\left(\left\{\Zb\left(\betab + \Eb\Sb^\T(\Sb\Sb^\T)^{-1}\right)\sbb_t\right\}^\T\epsilonb_{t+1}\right)_{t\in\Tc_1}\right]\right], \\
    & = \Ex\left[\left(\Zb\betab\sbb_t\right)^\T\betab\sbb_t\right]+\Ex\left[\Exs\left[\left\{\left(\Zb\Eb\Sb^\T(\Sb\Sb^\T)^{-1}\sbb_t\right)^\T \betab\sbb_t\right\}_{t\in\Tc_1}\right]\right] \\
    &+ \Ex\left[\Exs\left[\left(\left\{\Zb\Eb\Sb^\T(\Sb\Sb^\T)^{-1}\sbb_t\right\}^\T\epsilonb_{t+1}\right)_{t\in\Tc_1}\right]\right],
\end{align*}
since the sample mean is an unbiased estimator, and since \((\sbb_t)_{t}\) and \((\epsilonb_{t+1})_{t}\) are respectively iid, and \(\Ex[\epsilonb_{t+1}]=\zerob\).

Since we can ``extract'' the \(t\)-th column from a matrix \(\Xb\) by multiplying it by a vector~\(\eb_t\) which has \(1\) in the \(t\)-th position and zero everywhere else, then
\begin{align*}
    \Ex\left[\Exs\left[\left(\left(\Zb\Eb\Sb^\T(\Sb\Sb^\T)^{-1}\sbb_t\right)^\T \betab\sbb_t\right)_{t\in\Tc_1}\right]\right] &= \Ex\left[\eb_t^\T\Sb^\T(\Sb\Sb^\T)^{-1}\Sb\Eb^\T\Zb\betab\Sb\eb_t\right] = 0,\quad \text{for all } t \in \Tc_1,
\end{align*}
since $\Eb$ has a Matrix Normal distribution $\MN(\zerob,\Sigmab_\epsilon,\Id_{T_1})$ by assumption and since~$\Zb$ is assumed symmetric. As previously mentioned, the first equality is true for all \(t\in\Tc_1\) by the iid property of~\((\sbb_t)_{t}\), and the fact the sample mean is an unbiased estimator.

The next term reads
\begin{align*}
    \Ex\left[\Exs\left[\left(\left(\Zb\Eb\Sb^\T(\Sb\Sb^\T)^{-1}\sbb_t\right)^\T\epsilonb_{t+1}\right)_{t\in\Tc_1}\right]\right] &= \Ex\left[\eb_t^\T\Sb^\T(\Sb\Sb^\T)^{-1}\Sb\Eb^\T\Zb\Eb\eb_t\right],\quad \text{for any } t \in \Tc_1, \\
    &= \eb_t^\T\Ex[\Sb^\T(\Sb\Sb^\T)^{-1}\Sb]\Ex[\Eb^\T\Zb\Eb]\eb_t, \\
    &= \tr(\Zb\Sigmab_\epsilon)\eb_t^\T\Ex[\Sb^\T(\Sb\Sb^\T)^{-1}\Sb]\eb_t = \frac{p}{T_1}\tr(\Zb\Sigmab_\epsilon),
\end{align*}
where we have used the Matrix Normal identity
\begin{equation*}
    \Ex[\Xb\Ab\Xb^\T] = \Ub\tr(\Ab^\T\Vb)+\Mb\Ab\Mb^\T,
\end{equation*}
with \(\Xb\sim\MN(\Mb,\Ub,\Vb)\)~\cite[Theorem~2.3.5]{guptaMatrixVariateDistributions2000}.
We also used that \(\tr(\Ex[\Sb^\T(\Sb\Sb^\T)^{-1}\Sb])=p\) since the projection matrix $\Sb^\T(\Sb\Sb^\T)^{-1}\Sb$ is idempotent and thus its trace equals its rank, 
and since~\((\sbb_t)_{t}\) are iid, then the expectation of each element of the diagonal
of \(\Sb^\T(\Sb\Sb^\T)^{-1}\Sb\) are equal, thus \(\eb_t^\T\Ex[\Sb^\T(\Sb\Sb^\T)^{-1}\Sb]\eb_t=\frac{p}{T_1}\). Therefore, the in-sample expected P\&L is given by
\begin{equation*}
    \Ex\left[\Exs\left[(\widehat{\PnL}_t)_{t\in\Tc_1}\right]\right] = \tr\left(\betab\Sigmab_s\betab^\T\Zb\right) + \mub_s^\T\betab^\T\Zb\betab\mub_s + \frac{p}{T_1}\tr\left(\Zb\Sigmab_\epsilon\right).
\end{equation*}

\subsection{In-sample expected variance}
We turn next to the expected variance \(\Ex[\Vars[(\widehat{\PnL}_t)_{t\in\Tc_1}]]\):
\begin{align*}
    &\Ex\left[\Vars\left[\left((\Zb\widehat{\betab}\sbb_t)^\T(\betab\sbb_t+\epsilonb_{t+1})\right)_{t\in\Tc_1}\right]\right] \\
    &= \Ex\left[\Vars\left[\left(\left(\Zb\left(\betab + \Eb\Sb^\T(\Sb\Sb^\T)^{-1}\right)\sbb_t\right)^\T\betab\sbb_t+\left(\Zb\left(\betab + \Eb\Sb^\T(\Sb\Sb^\T)^{-1}\right)\sbb_t\right)^\T\epsilonb_{t+1}\right)_{t\in\Tc_1}\right]\right], \\
    &= \Ex\left[\Vars\left[\left((\Zb\betab\sbb_t)^\T\betab\sbb_t + (\Zb\Eb\Sb^\T(\Sb\Sb^\T)^{-1}\sbb_t)^\T\betab\sbb_t + (\Zb\betab\sbb_t)^\T\epsilonb_{t+1} + (\Zb\Eb\Sb^\T(\Sb\Sb^\T)^{-1}\sbb_t)^\T\epsilonb_{t+1}\right)_{t\in\Tc_1}\right]\right], \\
    &= \Ex\left[\Vars\left[\left((\Zb\betab\sbb_t)^\T\betab\sbb_t\right)_{t\in\Tc_1}\right]\right] + \Ex\left[\Vars\left[\left((\Zb\betab\sbb_t)^\T\epsilonb_{t+1}\right)_{t\in\Tc_1}\right]\right] + \Ex\left[\Vars\left[\left((\Zb\Eb\Sb^\T(\Sb\Sb^\T)^{-1}\sbb_t)^\T\betab\sbb_t\right)_{t\in\Tc_1}\right]\right] \\
    &+ \Ex\left[\Vars\left[\left((\Zb\Eb\Sb^\T(\Sb\Sb^\T)^{-1}\sbb_t)^\T\epsilonb_{t+1}\right)_{t\in\Tc_1}\right]\right] + \Ex\left[2\Covs\left[\left((\Zb\betab\sbb_t)^\T\betab\sbb_t\right)_{t\in\Tc_1},\left((\Zb\betab\sbb_t)^\T\epsilonb_{t+1}\right)_{t\in\Tc_1}\right]\right] \\
    &+ \Ex\left[2\Covs\left[\left((\Zb\betab\sbb_t)^\T\betab\sbb_t\right)_{t\in\Tc_1},\left((\Zb\Eb\Sb^\T(\Sb\Sb^\T)^{-1}\sbb_t)^\T\betab\sbb_t\right)_{t\in\Tc_1}\right]\right] \\
    &+ \Ex\left[2\Covs\left[\left((\Zb\betab\sbb_t)^\T\betab\sbb_t\right)_{t\in\Tc_1}, \left((\Zb\Eb\Sb^\T(\Sb\Sb^\T)^{-1}\sbb_t)^\T\epsilonb_{t+1}\right)_{t\in\Tc_1}\right]\right] \\
    &+ \Ex\left[2\Covs\left[\left((\Zb\betab\sbb_t)^\T\epsilonb_{t+1}\right)_{t\in\Tc_1}, \left((\Zb\Eb\Sb^\T(\Sb\Sb^\T)^{-1}\sbb_t)^\T\betab\sbb_t\right)_{t\in\Tc_1}\right]\right] \\
    &+ \Ex\left[2\Covs\left[\left((\Zb\betab\sbb_t)^\T\epsilonb_{t+1}\right)_{t\in\Tc_1},\left((\Zb\Eb\Sb^\T(\Sb\Sb^\T)^{-1}\sbb_t)^\T\epsilonb_{t+1}\right)_{t\in\Tc_1}\right]\right] \\
    &+ \Ex\left[2\Covs\left[\left((\Zb\Eb\Sb^\T(\Sb\Sb^\T)^{-1}\sbb_t)^\T\betab\sbb_t\right)_{t\in\Tc_1},\left((\Zb\Eb\Sb^\T(\Sb\Sb^\T)^{-1}\sbb_t)^\T\epsilonb_{t+1}\right)_{t\in\Tc_1}\right]\right].
\end{align*}
Since the signals and residuals are iid, and the sample variance is an unbiased estimator for the population variance, we can replace all the expected sample variance terms with the true variance. We are thus left with computing these true variances. The first two terms are
\begin{align*}
    \Ex\left[\Vars\left[\left((\Zb\betab\sbb_t)^\T\betab\sbb_t\right)_{t\in\Tc_1}\right]\right] &= \Var\left[\left(\Zb\betab\sbb_t\right)^\T\betab\sbb_t\right] = 2\tr\left(\Zb\betab \Sigmab_s \betab^\T \Zb \betab\Sigmab_s  \betab^\T\right) + 4\mub_s^\T\betab^\T\Zb \betab\Sigmab_s \betab^\T \Zb \betab \mub_s, \\
    \Ex\left[\Vars\left[\left((\Zb\betab\sbb_t)^\T\epsilonb_{t+1}\right)_{t\in\Tc_1}\right]\right] &= \Var\left[(\Zb\betab\sbb_t)^\T\epsilonb_{t+1}\right] = \tr\left(\Zb\betab\Sigmab_s\betab^\T\Zb\Sigmab_\epsilon\right) +\mub_s^\T\betab^\T\Zb\Sigmab_\epsilon\Zb\betab\mub_s,
\end{align*}
where the first equality in both expressions is for all \(t\in\Tc_1\), and the first equality comes from the iid assumption and the fact the sample variance is an unbiased estimator.

The next term is
\begin{align*}
    &\Ex\left[\Vars\left[\left((\Zb\Eb\Sb^\T(\Sb\Sb^\T)^{-1}\sbb_t)^\T\betab\sbb_t\right)_{t\in\Tc_1}\right]\right] = \Var\left[(\Zb\Eb\Sb^\T(\Sb\Sb^\T)^{-1}\sbb_t)^\T\betab\sbb_t\right],\quad \text{for all } t \in \Tc_1 \\
    &= \Ex\left[\left((\Zb\Eb\Sb^\T(\Sb\Sb^\T)^{-1}\sbb_t)^\T\betab\sbb_t\right)^2\right]-\Ex\left[\left(\Zb\Eb\Sb^\T(\Sb\Sb^\T)^{-1}\sbb_t\right)^\T\betab\sbb_t\right]^2 \\
    &= \Ex\left[\left((\Zb\Eb\Sb^\T(\Sb\Sb^\T)^{-1}\sbb_t)^\T\betab\sbb_t\right)^2\right],
\end{align*}
since $\Ex[\Eb]=0$ and~$\Eb$ and~$\Sb$ are independent.
To evaluate this further we again use 
\(\sbb_t=\Sb\eb_t\) 
and let 
\(\Hb\coloneqq\Sb^\T(\Sb\Sb^\T)^{-1}\Sb\) (the so called `hat' matrix in linear regression), so that
\begin{align*}
    \Ex\left[\left((\Zb\Eb\Sb^\T(\Sb\Sb^\T)^{-1}\sbb_t)^\T\betab\sbb_t\right)^2\right] &= \Ex\left[\left(\eb_t^\T\Sb^\T(\Sb\Sb^\T)^{-1}\Sb\Eb^\T\Zb\betab\Sb\eb_t\right)^2\right] \\
    &= \Ex\left[\eb_t^\T\Hb\Eb^\T\Zb\betab\Sb\eb_t\eb_t^\T\Hb\Eb^\T\Zb\betab\Sb\eb_t\right] \\
    &= \Ex\left[\eb_t^\T\Hb(\Id_{T_1}\Hb^\T\eb_t\eb_t^\T\Sb^\T\betab^\T\Zb\Sigmab_\epsilon)\Zb\betab\Sb\eb_t\right] \\
    &= \Ex\left[\eb_t^\T\Hb\eb_t\eb_t^\T\Sb^\T\betab^\T\Zb\Sigmab_\epsilon\Zb\betab\Sb\eb_t\right], \\
    &= \tr\left(\Ex\left[\eb_t^\T\Hb\eb_t\eb_t^\T\Sb^\T\betab^\T\Zb\Sigmab_\epsilon\Zb\betab\Sb\eb_t\right]\right) \\
    &= \tr\left(\Ex\left[(\Sb\Sb^\T)^{-1}\Sb\eb_t\eb_t^\T\Sb^\T\betab^\T\Zb\Sigmab_\epsilon\Zb\betab\Sb\eb_t\eb_t^\T\Sb^\T\right]\right) \\
    &= \tr\left(\Ex\left[(\Sb\Sb^\T)^{-1}\right]\Ex\left[\Sb\eb_t\eb_t^\T\Sb^\T\betab^\T\Zb\Sigmab_\epsilon\Zb\betab\Sb\eb_t\eb_t^\T\Sb^\T\right]+\mathcal{E}_1\right),
\end{align*}
where 
\begin{align*}
    \mathcal{E}_1 := \Ex\left[(\Sb\Sb^\T)^{-1}\Sb\eb_t\eb_t^\T\Sb^\T\betab^\T\Zb\Sigmab_\epsilon\Zb\betab\Sb\eb_t\eb_t^\T\Sb^\T\right]-\Ex\left[(\Sb\Sb^\T)^{-1}\right]\Ex\left[\Sb\eb_t\eb_t^\T\Sb^\T\betab^\T\Zb\Sigmab_\epsilon\Zb\betab\Sb\eb_t\eb_t^\T\Sb^\T\right].
\end{align*}
The \(\mathcal{E}_1\) term results from the fact that \((\Sb\Sb^\T)^{-1}\) is of course not independent of the other terms, however experimentally we find that \(\tr(\mathcal{E}_1)\) is negligible compared to the trace of the expectation product and decreases rapidly in the ratio \(p/T_1\). Provided we are in the regime where \(T_1 \gg p\), then this approximation is reasonable as can be seen in \Cref{fig:vareps_pct}.

With this, we can finish the evaluation. 
To simplify the expression it is convenient to swap~\(\Sb\eb_t\) with~\(\sbb_t\) again, and define the matrix \(\Fb\coloneqq\betab^\T\Zb\Sigmab_\epsilon\Zb\betab\), so that
\begin{align*} 
    &\Ex\left[\left((\Zb\Eb\Sb^\T(\Sb\Sb^\T)^{-1}\sbb_t)^\T\betab\sbb_t\right)^2\right] \\
    &= \tr\left(\Ex\left[(\Sb\Sb^\T)^{-1}\right]\Ex\left[\sbb_t\sbb_t^\T\betab^\T\Zb\Sigmab_\epsilon\Zb\betab\sbb_t\sbb_t^\T\right]+\mathcal{E}_1\right) \\
    &= \tr\left[\frac{\left(\Sigmab_s+\mub_s\mub_s^\T\right)^{-1}}{T_1-p-1}\left(2\left(\Sigmab_s+\mub_s\mub_s^\T\right)\Fb\left(\Sigmab_s+\mub_s\mub_s^\T\right)
    +\mub_s^\T\Fb\mub_s\left(\Sigmab_s-\mub_s\mub_s^\T\right)
    +\tr(\Fb\Sigmab_s)\left(\Sigmab_s+\mub_s\mub_s^\T\right)\right)\right] + \varepsilon_1 \\
    &= \frac{1}{T_1-p-1}\tr\left(2\Fb\left(\Sigmab_s+\mub_s\mub_s^\T\right)+\mub_s^\T\Fb\mub_s\cdot\left(\Sigmab_s+\mub_s\mub_s^\T\right)^{-1}\left(\Sigmab_s-\mub_s\mub_s^\T\right)+p\Fb\Sigmab_s\right) + \varepsilon_1,
\end{align*}
with \(\varepsilon_1 := \tr(\mathcal{E}_1)\). Thus, the expected variance term is given by
\begin{align*}
    &\Ex\left[\Vars\left[\left((\Zb\Eb\Sb^\T(\Sb\Sb^\T)^{-1}\sbb_t)^\T\betab\sbb_t\right)_{t\in\Tc_1}\right]\right] \\
    &= \frac{1}{T_1-p-1}\tr\left(2\Fb\left(\Sigmab_s+\mub_s\mub_s^\T\right)+\mub_s^\T\Fb\mub_s\cdot\left(\Sigmab_s+\mub_s\mub_s^\T\right)^{-1}\left(\Sigmab_s-\mub_s\mub_s^\T\right)+p\Fb\Sigmab_s\right) + \varepsilon_1.
\end{align*}

Moving on, we tackle the next term in the expected variance similarly:
\begin{align*}
    &\Ex\left[\Vars\left[\left((\Zb\Eb\Sb^\T(\Sb\Sb^\T)^{-1}\sbb_t)^\T\epsilonb_{t+1}\right)_{t\in\Tc_1}\right]\right] = \Var\left[(\Zb\Eb\Sb^\T(\Sb\Sb^\T)^{-1}\sbb_t)^\T\epsilonb_{t+1}\right],\quad \text{for all } t \in \Tc_1, \\
    &= \Ex\left[\left((\Zb\Eb\Sb^\T(\Sb\Sb^\T)^{-1}\sbb_t)^\T\epsilonb_{t+1}\right)^2\right]-\Ex\left[(\Zb\Eb\Sb^\T(\Sb\Sb^\T)^{-1}\sbb_t)^\T\epsilonb_{t+1}\right]^2 \\
    &= \Ex\left[\left((\Zb\Eb\Sb^\T(\Sb\Sb^\T)^{-1}\sbb_t)^\T\epsilonb_{t+1}\right)^2\right]-\left(\tr(\Zb\Sigmab_\epsilon)\frac{p}{T_1}\right)^2.
\end{align*}
We can evaluate the first term as
\begin{align*}
    &\Ex\left[\left((\Zb\Eb\Sb^\T(\Sb\Sb^\T)^{-1}\sbb_t)^\T\epsilonb_{t+1}\right)^2\right] \\
    &= \Ex\left[\eb_t\Hb\Eb^\T\Zb\Eb\eb_t\eb_t^\T\Eb^\T\Zb\Eb\Hb^\T\eb_t\right] \\
    &= \tr(\Sigmab_\epsilon\Zb\Sigmab_\epsilon\Zb)\left(\Ex\left[\eb_t^\T\Hb\eb_t\right]+\Ex\left[\eb_t^\T\Hb\eb_t\eb_t^\T\Hb^\T\eb_t\right]\right)+\tr(\Sigmab_\epsilon\Zb)^2\Ex\left[\eb_t^\T\Hb\eb_t\eb_t^\T\Hb\eb_t\right] \\
    &= \tr(\Sigmab_\epsilon\Zb\Sigmab_\epsilon\Zb)\left(\frac{p}{T_1}+\left(\frac{p(p+2)}{T_1(T_1+2)}-\uv_p^\T\mub_s\frac{2(T_1-p)(T_1-p+2)}{T^2(T+1)(T+2)}\right)\right)\\
    &+\tr(\Sigmab_\epsilon\Zb)^2\left(\frac{p(p+2)}{T_1(T_1+2)}-\uv_p^\T\mub_s\frac{2(T_1-p)(T_1-p+2)}{T^2(T+1)(T+2)}\right),
\end{align*}
where we have used the fact that when there is no intercept term the diagonal entries of the hat matrix \(\Hb\) are distributed \(\beta\left(\frac{p}{2}, \frac{(T-p)}{2}\right)\) \cite[Appendix 2A]{belsleyRegressionDiagnosticsIdentifying2004}, and for the case where there is an intercept term, \citeauthor{chatterjeeSensitivityAnalysisLinear1988}~\cite[Section~2.3.7]{chatterjeeSensitivityAnalysisLinear1988} derive an approximate distribution for the diagonal elements of the hat matrix,
\begin{align*}
    \frac{T_1-p}{p-1}\frac{h_{tt}-\frac{1}{T_1}}{1-h_{tt}} \dot\sim F(p-1,T_1-p),
\end{align*}
where \(F(p-1, T_1-p)\) denotes the \(F\) distribution with \(p-1\) and \(T_1-p\) degrees of freedom. However, according to \citeauthor{raoLinearStatisticalInference2009}~\cite[p. 542]{raoLinearStatisticalInference2009} this is an exact distribution, rather than an approximate one when \(\mu_s=\zerob\), which holds in our assumptions. With a little work this yields
\begin{align*}
    \Ex[h_{tt}^2] = \frac{p(p+2)}{T_1(T_1+2)}-\frac{2(T_1-p)(T_1-p+2)}{T^2(T+1)(T+2)},
\end{align*}
and noting that \(\uv_p^\T\mub_s\) is zero with no intercept, and one with an intercept, this yields the final line of the derivation above.
Putting it all together we obtain the expected variance term 
\begin{align*}
    &\Ex\left[\Vars\left[\left((\Zb\Eb\Sb^\T(\Sb\Sb^\T)^{-1}\sbb_t)^\T\epsilonb_{t+1}\right)_{t\in\Tc_1}\right]\right] \\
    &= \tr(\Sigmab_\epsilon\Zb\Sigmab_\epsilon\Zb)\left(\frac{p}{T_1}+\left(\frac{p(p+2)}{T_1(T_1+2)}-\uv_p^\T\mub_s\frac{2(T_1-p)(T_1-p+2)}{T^2(T+1)(T+2)}\right)\right)\\
    &+\tr(\Sigmab_\epsilon\Zb)^2\left(\frac{p(p+2)}{T_1(T_1+2)}-\uv_p^\T\mub_s\frac{2(T_1-p)(T_1-p+2)}{T^2(T+1)(T+2)}\right)-\left(\tr(\Zb\Sigmab_\epsilon)\frac{p}{T_1}\right)^2 \\
    &= \tr(\Sigmab_\epsilon\Zb\Sigmab_\epsilon\Zb)\frac{p T_1 \left(T_1+1\right) \left(p+T_1+4\right)-2 \left(p-T_1-2\right) \left(p-T_1\right) \uv_p^\T\mub_s}{T_1^2 \left(T_1+1\right)
    \left(T_1+2\right)} \\
    &-\tr(\Sigmab_\epsilon\Zb)^2\frac{2 \left(p-T_1\right) \left((p-2) \uv_p^\T\mub_s+T_1 \left(p-\uv_p^\T\mub_s\right)+p\right)}{T_1^2 \left(T_1+1\right) \left(T_1+2\right)}.\\
\end{align*}

Next we begin to tackle the covariance terms, again we use the fact that the sample covariance is an unbiased estimator and the iid property. The first two covariance terms are zero:
\begin{equation*}
    \Ex\left[2\Covs\left[\left((\Zb\betab\sbb_t)^\T\betab\sbb_t\right)_{t\in\Tc_1},\left((\Zb\betab\sbb_t)^\T\epsilonb_{t+1}\right)_{t\in\Tc_1}\right]\right] = 2\Cov[(\Zb\betab\sbb_t)^\T\betab\sbb_t,(\Zb\betab\sbb_t)^\T\epsilonb_{t+1}] = 0,
\end{equation*}
since \(\Ex[\epsilonb_{t+1}]=\zerob\), and
\begin{align*} 
    \Ex\left[2\Covs\left[\left((\Zb\betab\sbb_t)^\T\betab\sbb_t\right)_{t\in\Tc_1},\left((\Zb\Eb\Sb^\T(\Sb\Sb^\T)^{-1}\sbb_t)^\T\betab\sbb_t\right)_{t\in\Tc_1}\right]\right] &= 2\Cov\left[(\Zb\betab\sbb_t)^\T\betab\sbb_t,(\Zb\Eb\Sb^\T(\Sb\Sb^\T)^{-1}\sbb_t)^\T\betab\sbb_t\right]
\end{align*}
is null for all \(t\in\Tc_1\) as well since \(\Ex[\epsilonb_{t+1}]=\zerob\). 
Regarding the next covariance term,
\begin{align*}
    &\Ex\left[2\Covs\left[\left((\Zb\betab\sbb_t)^\T\betab\sbb_t\right)_{t\in\Tc_1}, \left((\Zb\Eb\Sb^\T(\Sb\Sb^\T)^{-1}\sbb_t)^\T\epsilonb_{t+1}\right)_{t\in\Tc_1}\right]\right] \\
    &= 2\Cov\left[(\Zb\betab\sbb_t)^\T\betab\sbb_t, (\Zb\Eb\Sb^\T(\Sb\Sb^\T)^{-1}\sbb_t)^\T\epsilonb_{t+1}\right],\quad \text{for all } t \in \Tc_1, \\
    &= 2\left(\Ex\left[(\Zb\betab\sbb_t)^\T\betab\sbb_t (\Zb\Eb\Sb^\T(\Sb\Sb^\T)^{-1}\sbb_t)^\T\epsilonb_{t+1}\right]-\Ex\left[(\Zb\betab\sbb_t)^\T\betab\sbb_t\right]\Ex\left[(\Zb\Eb\Sb^\T(\Sb\Sb^\T)^{-1}\sbb_t)^\T\epsilonb_{t+1}\right]\right).
\end{align*}
Starting with the first expectation,
\begin{align*}
    &\Ex\left[(\Zb\betab\sbb_t)^\T\betab\sbb_t (\Zb\Eb\Sb^\T(\Sb\Sb^\T)^{-1}\sbb_t)^\T\epsilonb_{t+1}\right] \\
    &= \Ex\left[\eb_t^\T\Sb^\T\betab^\T\Zb\betab\Sb\eb_t\eb_t^\T\Hb\Eb^\T\Zb\Eb\eb_t\right] \\
    &= \tr(\Zb\Sigmab_\epsilon)\Ex\left[\eb_t^\T\Sb^\T\betab^\T\Zb\betab\Sb\eb_t\eb_t^\T\Hb\eb_t\right] \\
    &= \tr(\Zb\Sigmab_\epsilon)\tr\left(\Ex\left[(\Sb\Sb^\T)^{-1}\right]\Ex\left[\sbb_t\sbb_t^\T\betab^\T\Zb\betab\sbb_t\sbb_t^\T\right]+\mathcal{E}_2\right) \\
    &= \frac{\tr(\Zb\Sigmab_\epsilon)\tr\left(2\betab^\T\Zb\betab(\Sigmab_s+\mub_s\mub_s^\T)+\mub_s^\T\betab^\T\Zb\betab\mub_s(\Sigmab_s+\mub_s\mub_s^\T)^{-1}(\Sigmab_s-\mub_s\mub_s^\T)+p\betab^\T\Zb\betab\Sigmab_s\right)}{T_1-p-1} + \varepsilon_2,
\end{align*}
where we have proceeded as before with a mixed expectation of \((\Sb\Sb^\T)^{-1}\) and \(\Sb\), and

\begin{equation*}
    \mathcal{E}_2 = \Ex\left[(\Sb\Sb^\T)^{-1}\sbb_t\sbb_t^\T\betab^\T\Zb\betab\sbb_t\sbb_t^\T\right]-\Ex\left[(\Sb\Sb^\T)^{-1}\right]\Ex\left[\sbb_t\sbb_t^\T\betab^\T\Zb\betab\sbb_t\sbb_t^\T\right],
\end{equation*}
and \(\varepsilon_2:=\tr(\mathcal{E}_2)\).
Next, the second expectation reads
\begin{equation*}
    \Ex\left[(\Zb\betab\sbb_t)^\T\betab\sbb_t\right]\Ex\left[(\Zb\Eb\Sb^\T(\Sb\Sb^\T)^{-1}\sbb_t)^\T\epsilonb_{t+1}\right] = \left(\tr(\betab\Sigmab_s\betab^\T\Zb)+\mub_s^\T\betab^\T\Zb\betab\mub_s\right)\frac{p}{T_1}\tr(\Zb\Sigmab_\epsilon).
\end{equation*}
Putting this together yields
\begin{align*}
    &\Ex\left[2\Covs\left[\left((\Zb\betab\sbb_t)^\T\betab\sbb_t\right)_{t\in\Tc_1}, \left((\Zb\Eb\Sb^\T(\Sb\Sb^\T)^{-1}\sbb_t)^\T\epsilonb_{t+1}\right)_{t\in\Tc_1}\right]\right]  \\
    &= \frac{2}{T_1-p-1}\tr(\Zb\Sigmab_\epsilon)\tr\left(2\betab^\T\Zb\betab(\Sigmab_s+\mub_s\mub_s^\T)+\mub_s^\T\betab^\T\Zb\betab\mub_s\cdot(\Sigmab_s+\mub_s\mub_s^\T)^{-1}(\Sigmab_s-\mub_s\mub_s^\T)+p\betab^\T\Zb\betab\Sigmab_s\right) \\
    &- \frac{2p}{T_1}\tr(\Zb\Sigmab_\epsilon)\left(\tr(\betab\Sigmab_s\betab^\T\Zb)+\mub_s^\T\betab^\T\Zb\betab\mub_s\right) + \varepsilon_2.
\end{align*}
The next covariance term is similar, namely
\begin{align*}
    &\Ex\left[2\Covs\left[\left((\Zb\betab\sbb_t)^\T\epsilonb_{t+1}\right)_{t\in\Tc_1}, \left((\Zb\Eb\Sb^\T(\Sb\Sb^\T)^{-1}\sbb_t)^\T\betab\sbb_t\right)_{t\in\Tc_1}\right]\right] \\
    &= 2\Cov\left[(\Zb\betab\sbb_t)^\T\epsilonb_{t+1}, (\Zb\Eb\Sb^\T(\Sb\Sb^\T)^{-1}\sbb_t)^\T\betab\sbb_t\right],\quad \text{for all } t \in \Tc_1, \\
    &= 2 \left(\Ex\left[(\Zb\betab\sbb_t)^\T\epsilonb_{t+1}(\Zb\Eb\Sb^\T(\Sb\Sb^\T)^{-1}\sbb_t)^\T\betab\sbb_t\right]-\Ex\left[(\Zb\betab\sbb_t)^\T\epsilonb_{t+1}\right]\Ex\left[(\Zb\Eb\Sb^\T(\Sb\Sb^\T)^{-1}\sbb_t)^\T\betab\sbb_t\right]\right) \\
    &= 2 \Ex\left[(\Zb\betab\sbb_t)^\T\epsilonb_{t+1}(\Zb\Eb\Sb^\T(\Sb\Sb^\T)^{-1}\sbb_t)^\T\betab\sbb_t\right],
\end{align*}
since \(\Ex[\epsilonb_{t+1}]=\zerob\).
The first expectation can be evaluated as
\begin{align*}
    &\Ex\left[(\Zb\betab\sbb_t)^\T\epsilonb_{t+1}(\Zb\Eb\Sb^\T(\Sb\Sb^\T)^{-1}\sbb_t)^\T\betab\sbb_t\right] \\
    &= \Ex\left[\eb_t^\T\Sb^\T\betab^\T\Zb\tr(\eb_t^\T\Sb^\T(\Sb\Sb^\T)^{-1}\Sb\eb_t)\Sigmab_\epsilon\Zb\betab\Sb\eb_t\right] \\
    &= \tr\left(\Ex\left[(\Sb\Sb^\T)^{-1}\right]\Ex\left[\sbb_t\sbb_t^\T\betab^\T\Zb\Zb\Sigmab_\epsilon\Zb\T\betab\sbb_t\sbb_t^\T\right]+\mathcal{E}_3\right) \\
    &= \frac{1}{T_1-p-1}\tr\left(2\Fb(\Sigmab_s+\mub_s\mub_s^\T)+\mub_s^\T\Fb\mub_s\cdot(\Sigmab_s+\mub_s\mub_s^\T)^{-1}(\Sigmab_s-\mub_s\mub_s^\T)+p\Fb\Sigmab_s\right) + \varepsilon_3,
\end{align*}
where again we handle the \((\Sb\Sb^\T)^{-1}\) terms as before, and
\begin{align*}
    \mathcal{E}_3 = \Ex\left[(\Sb\Sb^\T)^{-1}\sbb_t\sbb_t^\T\betab^\T\Zb\Zb\Sigmab_\epsilon\Zb\T\betab\sbb_t\sbb_t^\T\right]-\Ex\left[(\Sb\Sb^\T)^{-1}\right]\Ex\left[\sbb_t\sbb_t^\T\betab^\T\Zb\Zb\Sigmab_\epsilon\Zb\T\betab\sbb_t\sbb_t^\T\right],
\end{align*}
and \(\varepsilon_3:=\tr(\mathcal{E}_3)\). Thus,
\begin{align*}
    &\Ex\left[2\Covs\left[\left((\Zb\betab\sbb_t)^\T\epsilonb_{t+1}\right)_{t\in\Tc_1}, \left((\Zb\Eb\Sb^\T(\Sb\Sb^\T)^{-1}\sbb_t)^\T\betab\sbb_t\right)_{t\in\Tc_1}\right]\right] \\
    &= \frac{2}{T_1-p-1}\tr\left(2\Fb(\Sigmab_s+\mub_s\mub_s^\T)+\mub_s^\T\Fb\mub_s\cdot(\Sigmab_s+\mub_s\mub_s^\T)^{-1}(\Sigmab_s-\mub_s\mub_s^\T)+p\Fb\Sigmab_s\right) + \varepsilon_3,
\end{align*}
where again $\Fb = \betab^\T\Zb\Sigmab_\epsilon\Zb\betab$.
We turn now to the penultimate term:
\begin{align*}
    &\Ex\left[2\Covs\left[\left((\Zb\betab\sbb_t)^\T\epsilonb_{t+1}\right)_{t\in\Tc_1},\left((\Zb\Eb\Sb^\T(\Sb\Sb^\T)^{-1}\sbb_t)^\T\epsilonb_{t+1}\right)_{t\in\Tc_1}\right]\right] \\
    &= 2\Cov\left[(\Zb\betab\sbb_t)^\T\epsilonb_{t+1},(\Zb\Eb\Sb^\T(\Sb\Sb^\T)^{-1}\sbb_t)^\T\epsilonb_{t+1}\right],\quad \text{for all } t \in \Tc_1, \\
    &= 2\left(\Ex\left[(\Zb\betab\sbb_t)^\T\epsilonb_{t+1}(\Zb\Eb\Sb^\T(\Sb\Sb^\T)^{-1}\sbb_t)^\T\epsilonb_{t+1}\right]-\Ex\left[(\Zb\betab\sbb_t)^\T\epsilonb_{t+1}\right]\Ex\left[(\Zb\Eb\Sb^\T(\Sb\Sb^\T)^{-1}\sbb_t)^\T\epsilonb_{t+1}\right]\right) = 0,
\end{align*}
as the derivation yields a cubic in \(\Eb\) and \(\Ex[\epsilonb_{t+1}]=\zerob\).

For the final term, we can write
\begin{align*}
    &\Ex\left[2\Covs\left[\left((\Zb\Eb\Sb^\T(\Sb\Sb^\T)^{-1}\sbb_t)^\T\betab\sbb_t\right)_{t\in\Tc_1},\left((\Zb\Eb\Sb^\T(\Sb\Sb^\T)^{-1}\sbb_t)^\T\epsilonb_{t+1}\right)_{t\in\Tc_1}\right]\right] \\
    &= 2\Ex\left[\Exs\left[\left((\Zb\Eb\Sb^\T(\Sb\Sb^\T)^{-1}\sbb_t)^\T\betab\sbb_t(\Zb\Eb\Sb^\T(\Sb\Sb^\T)^{-1}\sbb_t)^\T\epsilonb_{t+1}\right)_{t\in\Tc_1}\right] \right.\\
    &\left.-\Exs\left[\left((\Zb\Eb\Sb^\T(\Sb\Sb^\T)^{-1}\sbb_t)^\T\betab\sbb_t\right)_{t\in\Tc_1}\right]\Exs\left[\left((\Zb\Eb\Sb^\T(\Sb\Sb^\T)^{-1}\sbb_t)^\T\epsilonb_{t+1}\right)_{t\in\Tc_1}\right]\right] =  0,
\end{align*}
which again yields terms with only \(\Eb\) or cubics in \(\Eb\), and from the identity for cubic expectations we see that every term in the expectation includes \(\mub_\epsilon\) which is zero, and thus the full expectation is zero. We can now collate all the terms for the expected variance \(\Ex[\Vars[(\widehat{\PnL}_t)_{t\in\Tc_1}]]\):
\begin{align*}   
    &\Ex[\Vars[(\widehat{\PnL}_t)_{t\in\Tc_1}]] \\
    &= 2\tr\left((\Gb\Sigmab_s)^2\right) + 4\mub_s^\T \Gb \Sigmab_s \Gb \mub_s
    + \tr(\Fb\Sigmab_s)+\mub_s^\T\Fb\mub_s \\
    &+ \frac{3}{T_1-p-1}\tr\left(2\Fb(\Sigmab_s+\mub_s\mub_s^\T)+\mub_s^\T\Fb\mub_s\cdot(\Sigmab_s+\mub_s\mub_s^\T)^{-1}(\Sigmab_s-\mub_s\mub_s^\T)+p\Fb\Sigmab_s\right) \\
    &+ \tr(\Sigmab_\epsilon\Zb\Sigmab_\epsilon\Zb)\frac{p T_1 \left(T_1+1\right) \left(p+T_1+4\right)-2 \left(p-T_1-2\right) \left(p-T_1\right) \uv_p^\T\mub_s}{T_1^2 \left(T_1+1\right)
    \left(T_1+2\right)} \\
    &-\tr(\Sigmab_\epsilon\Zb)^2\frac{2 \left(p-T_1\right) \left((p-2) \uv_p^\T\mub_s+T_1 \left(p-\uv_p^\T\mub_s\right)+p\right)}{T_1^2 \left(T_1+1\right) \left(T_1+2\right)}\\
    &+\frac{2}{T_1-p-1}\tr(\Zb\Sigmab_\epsilon)\tr\left(2\Gb(\Sigmab_s+\mub_s\mub_s^\T)+\mub_s^\T\Gb\mub_s\cdot(\Sigmab_s+\mub_s\mub_s^\T)^{-1}(\Sigmab_s-\mub_s\mub_s^\T)+p\Gb\Sigmab_s\right) \\
    &- \frac{2p}{T_1}\tr(\Zb\Sigmab_\epsilon)\left(\tr(\Gb\Sigmab_s)+\mub_s^\T\Gb\mub_s\right) \\
    &+ \varepsilon_1 + \varepsilon_2 + \varepsilon_3,
\end{align*}
where
\begin{align*}
\Fb := \betab^\T\Zb\Sigmab_\epsilon\Zb\betab
\qquad\text{and}\qquad
\Gb := \betab^\T\Zb\betab,
\end{align*}
and
\begin{align*}
    \varepsilon_1 &= \tr\left(\Ex\left[(\Sb\Sb^\T)^{-1}\Sb\eb_t\eb_t^\T\Sb^\T\betab^\T\Zb\Sigmab_\epsilon\Zb\betab\Sb\eb_t\eb_t^\T\Sb^\T\right]-\Ex\left[(\Sb\Sb^\T)^{-1}\right]\Ex\left[\Sb\eb_t\eb_t^\T\Sb^\T\betab^\T\Zb\Sigmab_\epsilon\Zb\betab\Sb\eb_t\eb_t^\T\Sb^\T\right]\right), \\
    \varepsilon_2 &= \tr\left(\Ex\left[(\Sb\Sb^\T)^{-1}\sbb_t\sbb_t^\T\betab^\T\Zb\betab\sbb_t\sbb_t^\T\right]-\Ex\left[(\Sb\Sb^\T)^{-1}\right]\Ex\left[\sbb_t\sbb_t^\T\betab^\T\Zb\betab\sbb_t\sbb_t^\T\right]\right), \\
    \varepsilon_3 &= \tr\left(\Ex\left[(\Sb\Sb^\T)^{-1}\sbb_t\sbb_t^\T\betab^\T\Zb\Zb\Sigmab_\epsilon\Zb\T\betab\sbb_t\sbb_t^\T\right]-\Ex\left[(\Sb\Sb^\T)^{-1}\right]\Ex\left[\sbb_t\sbb_t^\T\betab^\T\Zb\Zb\Sigmab_\epsilon\Zb\T\betab\sbb_t\sbb_t^\T\right]\right).
\end{align*}

\subsection{Out-of-sample expected average}
Next we derive the expected out-of-sample expected return and variance.
We now have that \(\widehat{\betab}\) is independent of the realisations of the signal and the residuals, 
thus the derivation is simplified significantly. Firstly, the expected out-of-sample P\&L is given by
\begin{equation*}    \Ex\left[\Exs\left[\left((\Zb\widehat{\betab}\sbb_t)^\T(\betab\sbb_t+\epsilonb_{t+1})\right)_{t\in\Tc_2}\right]\right] = \Ex[(\Zb\betab\sbb_t)^\T\betab\sbb_t] = \tr(\betab\Sigmab_s\betab^\T\Zb)+\mub_s^\T\betab^\T\Zb\betab\mub_s,
\end{equation*}
where we have no 
\(\Ex[(\Zb\betab\sbb_t)^\T\epsilonb_{t+1}]\) term since \(\Ex[\epsilonb_{t+1}]=\zerob\) and by independence.

\subsection{Out-of-sample expected variance}
Next the variance is given by
\begin{align*}
    &\Ex\left[\Vars\left[\left((\Zb\widehat{\betab}\sbb_t)^\T(\betab\sbb_t+\epsilonb_{t+1})\right)_{t\in\Tc_2}\right]\right] \\
    &= \Ex\left[\Vars\left[\left((\Zb\widehat{\betab}\sbb_t)^\T\betab\sbb_t\right)_{t\in\Tc_2}\right]+\Vars\left[\left((\Zb\widehat{\betab}\sbb_t)^\T\epsilonb_{t+1}\right)_{t\in\Tc_2}\right] + 2\Covs\left[\left((\Zb\widehat{\betab}\sbb_t)^\T\betab\sbb_t\right)_{t\in\Tc_2},\left((\Zb\widehat{\betab}\sbb_t)^\T\epsilonb_{t+1}\right)_{t\in\Tc_2}\right]\right].
\end{align*}

The first term can be evaluated as
\begin{align*}
    \Ex\left[\Vars\left[\left((\Zb\widehat{\betab}\sbb_t)^\T\betab\sbb_t\right)_{t\in\Tc_2}\right]\right] &= \Ex\left[\Vars\left[\left((\Zb\betab\sbb_t)^\T\betab\sbb_t\right)_{t\in\Tc_2}\right]+\Vars\left[\left((\Zb\Eb\Sb^\T(\Sb\Sb^\T)^{-1}\sbb_t)^\T\betab\sbb_t\right)_{t\in\Tc_2}\right]\right.\\
    &\left.+2\Covs\left[\left((\Zb\betab\sbb_t)^\T\betab\sbb_t\right)_{t\in\Tc_2},\left((\Zb\Eb\Sb^\T(\Sb\Sb^\T)^{-1}\sbb_t)^\T\betab\sbb_t\right)_{t\in\Tc_2}\right]\right],
\end{align*}
where we must be careful to note that \(\Eb\Sb^\T(\Sb\Sb^\T)^{-1}\), i.e.\ the error in \(\widehat{\betab}\), are the errors and signals from the first period, not the second. The first term of this is simply given by
\begin{equation*}\Ex\left[\Vars\left[\left((\Zb\betab\sbb_t)^\T\betab\sbb_t\right)_{t\in\Tc_2}\right]\right] = \Var[(\Zb\betab\sbb_t)^\T\betab\sbb_t] = 2\tr(\Zb\betab \Sigmab_s \betab^\T \Zb \betab\Sigmab_s  \betab^\T) + 4\mub_s^\T\betab^\T\Zb \betab\Sigmab_s \betab^\T \Zb \betab \mub_s,
\end{equation*}
where the first equality holds for all \(t\in\Tc_2\) since \(\sbb_t\) are iid. 
The covariance term is zero since 
\(\Ex[\epsilonb_{t+1}]=\zerob\), 
so we are left with \(\Ex\left[\Vars\left[\left((\Zb\Eb\Sb^\T(\Sb\Sb^\T)^{-1}\sbb_t)^\T\betab\sbb_t\right)_{t\in\Tc_2}\right]\right]\).
With \(\varepsilonb_\beta\coloneqq\Eb\Sb^\T(\Sb\Sb^\T)^{-1}\),
\begin{align*} &\Ex\left[\Vars\left[\left((\Zb\Eb\Sb^\T(\Sb\Sb^\T)^{-1}\sbb_t)^\T\betab\sbb_t\right)_{t\in\Tc_2}\right]\right] = \Var\left[(\Zb\varepsilonb_\beta\sbb_t)^\T\betab\sbb_t\right],\quad \text{for all } t \in \Tc_2,\\
    &= \Ex\left[\tr\left(\varepsilonb_\beta^\T\Zb\betab\Sigmab_s\left(\varepsilonb_\beta^\T\Zb\betab+\betab^\T\Zb\varepsilonb_\beta\right)\Sigmab_s\right)\right. \left.+\mub_s^\T\left(\varepsilonb_\beta^\T\Zb\betab+\betab^\T\Zb\varepsilonb_\beta\right)\Sigmab_s\left(\varepsilonb_\beta^\T\Zb\betab+\betab^\T\Zb\varepsilonb_\beta\right)\mub_s\right].
\end{align*}
To evaluate this we require some intermediate expectations,
\begin{align*}
    \Ex\left[\varepsilonb_\beta^\T\Zb\betab\Sigmab_s\varepsilonb_\beta^\T\Zb\right] &= \Db\Sigmab_s^\T\betab^\T\Zb\Sigmab_\epsilon\Zb, \\
    \Ex\left[\varepsilonb^\T\Zb\betab\Sigmab_s\betab^\T\Zb\varepsilonb\right] &= \Db\tr\left(\Sigmab_s\betab^\T\Zb\Sigmab_\epsilon\Zb\betab\right), \\
    \Ex\left[\varepsilonb_\beta\Sigmab_s\varepsilonb_\beta^\T\right] &= \Sigmab_\epsilon\tr\left(\Sigmab_s^\T\Db\right),
\end{align*}
where
\begin{equation*}
    \Db \coloneqq \frac{\left(\Sigmab_s+\mub_s\mub_s^\T\right)^{-1}}{T_1-p-1}.
\end{equation*}
Putting this all together yields
\begin{align*}
    &\Ex\left[\Vars\left[\left((\Zb\Eb\Sb^\T(\Sb\Sb^\T)^{-1}\sbb_t)^\T\betab\sbb_t\right)_{t\in\Tc_2}\right]\right] \\
    &= \tr\left(\Db\Sigmab_s^\T\betab^\T\Zb\Sigmab_\epsilon\Zb\betab\Sigmab_s\right)+\tr\left(\Db\Sigmab_s\right)\tr\left(\Sigmab_\epsilon\Zb\betab\Sigmab_s^\T\betab^\T\Zb\right)
    \\
    &+\mub_s^\T\left(2\Db\Sigmab_s^\T\betab^\T\Zb\Sigmab_\epsilon\Zb\betab+\tr\left(\Sigmab_\epsilon\Zb\betab\Sigmab_s^\T\betab^\T\Zb\right)\Db+\tr\left(\Sigmab_s^\T\Db\right)\betab^\T\Zb\Sigmab_\epsilon\Zb\betab\right)\mub_s.
\end{align*}
Thus, the first term for the expected variance out-of-sample is
\begin{align*}
    &\Ex\left[\Vars\left[\left((\Zb\widehat{\betab}\sbb_t)^\T\betab\sbb_t\right)_{t\in\Tc_2}\right]\right] \\
    &= 2\tr(\Zb\betab \Sigmab_s \betab^\T \Zb \betab\Sigmab_s  \betab^\T) + 4\mub_s^\T\betab^\T\Zb \betab\Sigmab_s \betab^\T \Zb \betab \mub_s \\
    &+ \tr\left(\Db\Sigmab_s^\T\betab^\T\Zb\Sigmab_\epsilon\Zb\betab\Sigmab_s\right)+\tr\left(\Db\Sigmab_s\right)\tr\left(\Sigmab_\epsilon\Zb\betab\Sigmab_s^\T\betab^\T\Zb\right)
    \\
    &+\mub_s^\T\left(2\Db\Sigmab_s^\T\betab^\T\Zb\Sigmab_\epsilon\Zb\betab+\tr\left(\Sigmab_\epsilon\Zb\betab\Sigmab_s^\T\betab^\T\Zb\right)\Db+\tr\left(\Sigmab_s^\T\Db\right)\betab^\T\Zb\Sigmab_\epsilon\Zb\betab\right)\mub_s.
\end{align*}
We can proceed similarly for the second term,
\begin{align*}
    \Ex\left[\Vars\left[\left((\Zb\widehat{\betab}\sbb_t)^\T\epsilonb_{t+1}\right)_{t\in\Tc_2}\right]\right] &= \Ex\left[\Vars\left[\left((\Zb\betab\sbb_t)^\T\epsilonb_{t+1}\right)_{t\in\Tc_2}\right]+\Vars\left[\left((\Zb\Eb\Sb^\T(\Sb\Sb^\T)^{-1}\sbb_t)^\T\epsilonb_{t+1}\right)_{t\in\Tc_2}\right]\right. \\
    &\left.+2\Covs\left[\left((\Zb\betab\sbb_t)^\T\epsilonb_{t+1}\right)_{t\in\Tc_2},\left((\Zb\Eb\Sb^\T(\Sb\Sb^\T)^{-1}\sbb_t)^\T\epsilonb_{t+1}\right)_{t\in\Tc_2}\right]\right].
\end{align*}
Again the first term is simply
\begin{align*}
    \Ex\left[\Vars\left[\left((\Zb\betab\sbb_t)^\T\epsilonb_{t+1}\right)_{t\in\Tc_2}\right]\right] &= \Var\left[(\Zb\betab\sbb_t)^\T\epsilonb_{t+1}\right] = \tr(\Zb\betab\Sigmab_s\betab^\T\Zb\Sigmab_\epsilon)+\mub_s^\T\betab^\T\Zb\Sigmab_\epsilon\Zb\betab\mub_s,
\end{align*}
and the final term is zero as \(\Eb\) is from the in-sample period, and therefore independent of \(\epsilonb_{t+1}\) from the out-of-sample period, and \(\Ex[\epsilonb_{t+1}]=\zerob\). Thus, we need only to find the second term, 
\begin{align*}
    \Ex\left[\Vars\left[\left((\Zb\Eb\Sb^\T(\Sb\Sb^\T)^{-1}\sbb_t)^\T\epsilonb_{t+1}\right)_{t\in\Tc_2}\right]\right] &= \Ex\left[\tr(\Zb\varepsilonb_\beta\Sigmab_s\varepsilonb_\beta^\T\Zb\Sigmab_\epsilon)+\mub_s^\T\varepsilonb_\beta^\T\Zb\Sigmab_\epsilon\Zb\varepsilonb_\beta\mub_s\right] \\
    &= \tr\left((\Zb\Sigmab_\epsilon)^2\right)\left(\tr(\Sigmab_s^\T\Db)+\mub_s^\T\Db\mub_s\right).
\end{align*}
And therefore, the full second term is given by
\begin{equation*}\Ex\left[\Vars\left[\left((\Zb\widehat{\betab}\sbb_t)^\T\epsilonb_{t+1}\right)_{t\in\Tc_2}\right]\right]
    =  \tr(\Zb\betab\Sigmab_s\betab^\T\Zb\Sigmab_\epsilon)+\mub_s^\T\betab^\T\Zb\Sigmab_\epsilon\Zb\betab\mub_s + \tr\left((\Zb\Sigmab_\epsilon)^2\right)\left(\tr(\Sigmab_s^\T\Db)+\mub_s^\T\Db\mub_s\right).
\end{equation*}
The third term, the covariance term, is necessarily zero by \(\Ex[\epsilonb_{t+1}]=\zerob\) and independence, thus the expected out-of-sample variance is given by
\begin{align*}
    &\Ex\left[\Vars\left[\left((\Zb\widehat{\betab}\sbb_t)^\T(\betab\sbb_t+\epsilonb_{t+1})\right)_{t\in\Tc_2}\right]\right] \\
    &= 2\tr(\Zb\betab \Sigmab_s \betab^\T \Zb \betab\Sigmab_s  \betab^\T) + 4\mub_s^\T\betab^\T\Zb \betab\Sigmab_s \betab^\T \Zb \betab \mub_s \\
    &+ \tr\left(\Db\Sigmab_s^\T\betab^\T\Zb\Sigmab_\epsilon\Zb\betab\Sigmab_s\right)+\tr\left(\Db\Sigmab_s\right)\tr\left(\Sigmab_\epsilon\Zb\betab\Sigmab_s^\T\betab^\T\Zb\right)
    \\
    &+\mub_s^\T\left(2\Db\Sigmab_s^\T\betab^\T\Zb\Sigmab_\epsilon\Zb\betab+\tr\left(\Sigmab_\epsilon\Zb\betab\Sigmab_s^\T\betab^\T\Zb\right)\Db+\tr\left(\Sigmab_s^\T\Db\right)\betab^\T\Zb\Sigmab_\epsilon\Zb\betab\right)\mub_s \\
    & +\tr(\Zb\betab\Sigmab_s\betab^\T\Zb\Sigmab_\epsilon)+\mub_s^\T\betab^\T\Zb\Sigmab_\epsilon\Zb\betab\mub_s + \tr\left((\Zb\Sigmab_\epsilon)^2\right)\left(\tr(\Sigmab_s^\T\Db)+\mub_s^\T\Db\mub_s\right).
\end{align*}

\section{Monte Carlo Details}\label{sec:mc_analysis}
As previously mentioned, we use Monte Carlo simulations both to check that the approximations~\eqref{eq:eoos_eis_approx} are reasonable (\(\varepsilon_i=0\) and ignoring the convexity adjustment) and to support and provide additional analysis. To do this, we observe the point estimates of the parameters from the experiments using commodity futures in \Cref{sec:pred_com_futures} and select distributions to draw new realisations within a cloud around these point estimates. Specifically,
\begin{itemize}
    \item For the signals we let the standard deviation of the signals be \(1\), which is unrestrictive as we can always standardise any given signal.
    \item For the residuals we sample the variances from a \(\chi^2\) distribution, which is well known to be the sample distribution for variance.
    \item For both signals and residuals we draw correlation matrices from the LJK distribution~\cite{lewandowskiGeneratingRandomCorrelation2009}, commonly used as a prior for correlation matrices in Bayesian modelling, and which can be considered the correlation equivalent of the Wishart distribution used for covariance matrices.
    \item We can then construct the covariance matrix \(\Sigmab=\textnormal{diag}(\sigmab)\mathbf{P}\textnormal{diag}(\sigmab)\), where~\(\sigmab\) is the vector of variances per signal/residual and \(\mathbf{P}\) is the sampled correlation matrix for the signal/residuals.
    \item For \(\betab\) we sample entries from a Laplace distribution, which provided the best fit when looking at the entries of real \(\betab\) matrices from \Cref{sec:pred_com_futures}.
\end{itemize}

With \(\betab\) and \(\Sigmab_\epsilon,\Sigmab_s\) coming from real data we would expect certain structures such as clusters with the covariance matrix (e.g.\ similarity between stocks in the same sector), or similar parameters for certain stocks;
this random sampling unfortunately misses out on this. However, from studying the expressions for the replication ratio it does not appear that these characteristics would impact our results. This is further confirmed by noting that when using real assets and signals in \Cref{sec:pred_com_futures}, the results remain valid. To achieve a desired true Sharpe ratio we simply scale the matrix~\(\betab\) appropriately. Finally, we let \(\mub_s=\zerob\) and do not include a drift term.

We note here that to simulate even 100 assets requires a significant amount of compute, particularly with large \(T\), which realistically will almost always be the case for a strategy of this style. Thus, the analytical approximations are useful to enable quick comparative adjustments for strategies, without requiring lengthy simulations. 

\section{Additional Values for \Cref{sec:pred_com_futures} and \Cref{sec:gwz}}

\subsection*{\Cref{sec:pred_com_futures} - Estimated Covariance}\label{sec:addnl_sec5}

In \Cref{sec:pred_com_futures} we have used the true covariance matrix when constructing our portfolios. To further confirm that the values are robust to the estimation of the covariance matrix (as previously shown in \Cref{sec:unknown_cov}), we run the same simulation again, but this time estimating the covariance matrix for each sample from the observed in-sample data, and using that to construct the portfolio \(\widehat{\wt}=\widehat{\Sigmab}_\epsilon^{-1}\widehat{\betab}\sbb_t\).

We collect all the metrics for the analytical, Gaussian + IID simulated estimates, and t + AR(1) simulated estimates into \Cref{tab:all_sim_metrics_futs}. Overall, we see that replacing the true residual covariance matrix with its in-sample estimate has only a negligible impact on both in-sample and out-of-sample performance, in line with the sensitivity analysis in \Cref{sec:unknown_cov}. Additionally, we see that the average realised Sharpe ratio is extremely close to the estimate which does not account for convexity (yielded by dividing the average mean by the square root of the average variance), which is in turn close to our analytical value. Thus, to conclude this section, we see that even when the data generating process departs from our idealised assumption, the analytical estimates are still a good estimator for all the metrics of interest.

\begin{table}[!ht]
    \centering
    \begin{tabular}{lccccc}
        \toprule
        & & \multicolumn{2}{c}{\textbf{Gaussian + IID}} & \multicolumn{2}{c}{\textbf{\(t\) + AR(1)}} \\
        \cmidrule(lr){3-4} \cmidrule(lr){5-6}
        & Analytical & True Cov & Est. Cov & True Cov & Est. Cov \\
        \midrule
        \multicolumn{6}{l}{\textit{In-Sample}} \\
        Mean $\mu$ & 0.28 & 0.28 & 0.29 & 0.31 & 0.32 \\
        Variance $\sigma^2$ & 0.30 & 0.30 & 0.30 & 0.41 & 0.38 \\
        Sharpe Ratio $\SR_\IS$ & N/A & 8.24 & 8.26 & 7.95 & 8.12 \\
        $\mu/\sigma$ ($\SR_\EIS$) & 8.22 & 8.24 & 8.26 & 7.72 & 8.10 \\
        \midrule
        \multicolumn{6}{l}{\textit{Out-of-Sample}} \\
        Mean $\mu$ & 0.11 & 0.11 & 0.11 & 0.14 & 0.14 \\
        Variance $\sigma^2$ & 0.29 & 0.29 & 0.30 & 0.59 & 0.62 \\
        Sharpe Ratio $\SR_\OOS$ & N/A & 3.14 & 3.13 & 2.83 & 2.83 \\
        $\mu/\sigma$ ($\SR_\EOOS$) & 3.13 & 3.14 & 3.13 & 2.79 & 2.79 \\
        \midrule
        \multicolumn{6}{l}{\textit{Replication Ratio}} \\
        $\SR_\OOS/\SR_\IS$ & N/A & 38\% & 38\% & 36\% & 35\% \\
        $\SR_\EOOS/\SR_\EIS$ & 38\% & 38\% & 38\% & 36\% & 34\% \\
        \bottomrule
    \end{tabular}
    \caption{Simulation Results comparing Analytical and Estimated values under Gaussian IID and AR(1)+t assumptions for \Cref{sec:pred_com_futures}. Sharpe ratios are annualised assuming 252 trading days.}\label{tab:all_sim_metrics_futs}
\end{table}

\subsection*{\Cref{sec:gwz} - Full Metrics}\label{sec:addnl_sec}
We report here the expected metrics in the same style as the previous section. Note that the results reported are averages over all 5{,}000{,}000 samples. For example, the interpretation of the empirical expected in-sample mean is: ``The average observed in-sample mean for all combinations of signals over all sub-periods'', and the interpretation of the analytical expected in-sample mean is: ``The average expected in-sample mean for all combinations of signals over all sub-periods''.

We observe that while the raw mean and variance estimates differ significantly between the analytical and empirical results, this discrepancy is primarily attributable to a scale factor. Most notably, the large empirical out-of-sample variance of $44.58$ (compared to the analytical $0.27$) is driven by significant kurtosis in the empirical values; for comparison, the modal values are $0.05$ and $0.09$ and the median values are $0.49$ and $0.12$ respectively. In particular, some resamples generate very large realised variances for portfolios whose corresponding out-of-sample Sharpe ratios are essentially zero. These extreme observations have little impact on the average Sharpe ratio but inflate the average variance. As the portfolio weights can be multiplied by a constant to scale both the mean and variance without altering the risk-adjusted return, the scale-invariant Sharpe ratio provides a more robust metric for comparison.

In this regard, the analytical and empirical estimates align remarkably well. The in-sample expected analytical Sharpe ratio $\Ex[\SR_\EIS]=1.16$ is comparable to the derived empirical ratio of~$1.21$. Similarly, the out-of-sample estimates are similar for the analytical and empirical values: the analytical model predicts $\Ex[\SR_\EOOS]=0.18$, which matches the average observed $\Ex[\SR_\OOS]=0.18$. This alignment results in consistent replication ratios across the different aggregation methods, ranging between $46\%$ and $54\%$.

\begin{table}[!ht]
    \centering
    \begin{tabular}{lcc}
        \toprule
        & \textbf{Analytical} & \textbf{Empirical} \\
        \midrule
        \multicolumn{3}{l}{\textit{In-Sample}} \\
        Mean $\Ex[\mu]$ & 0.19 & 0.33 \\
        Variance $\Ex[\sigma^2]$ & 0.32 & 0.87 \\
        Sharpe Ratio $\Ex[\SR_\IS]$ & N/A & 1.33 \\
        $\mu/\sigma$ ($\Ex[\SR_\EIS]$) & 1.16 & 1.21 \\
        \midrule
        \multicolumn{3}{l}{\textit{Out-of-Sample}} \\
        Mean $\Ex[\mu]$ & 0.09 & 0.41 \\
        Variance $\Ex[\sigma^2]$ & 0.27 & 44.58 \\
        Sharpe Ratio $\Ex[\SR_\OOS]$ & N/A & 0.18 \\
        $\mu/\sigma$ ($\Ex[\SR_\EOOS]$) & 0.18 & 0.21 \\
        \midrule
        \multicolumn{3}{l}{\textit{Replication Ratio}} \\
        $\Ex[\SR_\OOS/\SR_\IS]$ & N/A & 52\% \\
        $\Ex[\SR_\OOS]/\Ex[\SR_\IS]$ & N/A & 46\% \\
        $\Ex[\SR_\EOOS/\SR_\EIS]$ & 54\% & N/A \\
        $\Ex[\SR_\EOOS]/\Ex[\SR_\EIS]$ & 52\% & N/A \\
        \bottomrule
    \end{tabular}
    \caption{Statistics from resampling study, comparing analytical (implied) and empirical values for \Cref{sec:gwz}. Sharpe ratios are annualised from monthly values.}\label{tab:sec6_stats}
\end{table}

\section*{Disclaimer}
Unless specifically indicated otherwise, the views and opinions expressed in this communication are those of the author(s) and do not reflect the official policy or position of [Qube Research \& Technologies Ltd]. The information provided herein is for informational purposes only and should not be construed as advice, financial or otherwise, or opinion of [Qube Research \& Technologies Ltd]. [Qube Research \& Technologies Ltd] has neither approved nor disapproved of such content and makes no representations as to the accuracy, completeness, or reliability of any information contained herein.

Recipients are advised not to rely solely on this communication for any purpose. [Qube Research \& Technologies Ltd] shall not be liable for any errors or omissions, nor for any damages resulting from reliance on this information.

\end{document}